\documentclass{openjournal}
\usepackage{xcolor}
\usepackage{textgreek}
\usepackage[utf8]{inputenc}
\usepackage[english]{babel}
\usepackage{graphicx}	
\usepackage{amsmath}
\usepackage{threeparttable}
\usepackage{orcidlink}

\usepackage{hyperref}
\hypersetup{
    unicode, 
    colorlinks=true,
    linkcolor=linkcolor,
    citecolor=linkcolor,
    filecolor=linkcolor,
    urlcolor=linkcolor,
}
\usepackage{color,colortbl}
\definecolor{linkcolor}{rgb}{0.0,0.3,0.5}
\usepackage{tensind}
\tensordelimiter{?}
\DeclareGraphicsExtensions{.bmp,.png,.jpg,.pdf}
\usepackage[normalem]{ulem}
\usepackage{orcidlink}
\usepackage{soul}

\graphicspath{ {./figs/} }

\begin{document}

\title{Synchrotron Filaments in Radio Galaxies: Rarity, Stability, and Environmental Influences from LOFAR and \textit{ROSAT} Observations}

\author{Geoffrey Ong'alo\orcidlink{0009-0008-1860-559X}\altaffilmark{1,2},
Martin J. Hardcastle\orcidlink{0000-0003-4223-1117}\altaffilmark{2}, 
Willice Obonyo\orcidlink{0000-0001-9038-1756}\altaffilmark{3}, 
Marcus Brüggen\orcidlink{0000-0002-3369-7735}\altaffilmark{4}, 
and Paul Baki\altaffilmark{1}}

\email{geoffrey.ongalo@tukenya.ac.ke}

\affil{$^1$Department of Physics, Earth and Environmental Sciences, Technical University of Kenya, PO Box 52428-00200, Nairobi, Kenya}
\affil{$^2$Centre for Astrophysics Research, Department of Physics, Astronomy and Mathematics, University of Hertfordshire, College Lane, Hatfield AL10 9AB, UK}
\affil{$^3$Department of Mathematical Sciences, University of South Africa, Cnr Christian de Wet Rd and Pioneer Avenue, Florida Park, 1709, Roodepoort, South Africa.}
\affil{$^4$Hamburger Sternwart, University of Hamburg, Gojenbergsweg 112, 21029 Hamburg, Germany}

\begin{abstract}
We present a systematic investigation of filamentary synchrotron structures in radio galaxies, using deep 144 MHz observations from the LOFAR Two-Metre Sky Survey (LoTSS) in combination with archival \textit{ROSAT} X-ray data. These narrow, collimated features, linking radio lobes and often aligned with jet axes, are rare: our visual inspection of 548 cluster-associated sources identified 12 candidates ($\sim$2.2\%), of which three exhibit clearly resolved filaments suitable for detailed X-ray and synchrotron minimum pressure analysis. Minimum pressures in these filaments are consistently lower than the ambient thermal pressure derived from X-ray measurements. This disparity implies a role for external confinement in filament stability, aligning with theoretical predictions of thermal pressure support and potentially toroidal magnetic field configurations. We discuss filament formation scenarios, including relic AGN outflows, magnetic draping, and plasma instabilities. Our findings highlight the utility of filamentary structures as probes of intracluster magnetic field topology and AGN feedback, and demonstrate the need for high-resolution, multi-frequency studies to constrain their origin and evolution.
\end{abstract}


\maketitle

\section{Introduction}
\label{sec:intro}

The presence of filamentary synchrotron structures in radio galaxies has been a topic of increasing interest due to their implications for understanding the complex interactions between galactic jets, magnetic fields, and the intra-cluster medium (ICM). Early observations of radio galaxies typically highlighted large-scale features such as lobes and jets, with little focus on finer substructures. However, deep low-frequency observations are increasingly revealing a broader morphological class of macroscopic, spatially coherent synchrotron filaments that protrude from or connect radio lobes. A highly specific, extreme subclass of these features was revealed by MeerKAT observations of ESO 137$-$006, which identified features described as `collimated synchrotron threads' (CSTs) \citep{Ramatsoku2020} that extend for large distances \textit{outside} the radio lobes.

Abell 194 offers another case study in this context. This nearby, dynamically quiescent cluster hosts two prominent, distorted radio galaxies: the wide-angle tail (WAT) source 3C\,40B and the narrow-angle tail (NAT) source 3C\,40A, associated with NGC 547 and NGC 541, respectively. Both sources are located near the cluster core and exhibit bent morphologies commonly attributed to ram pressure interactions with the ICM \citep{Sakelliou2008}. Both earlier work \citep{Sakelliou2008} and more recent studies \citep{Rudnick2022} have reported the presence of narrow, elongated synchrotron structures in Abell 194, interpreted as signatures of interactions between AGN jets and magnetised filaments embedded in the ICM. These features, observed in the absence of strong merger activity, highlight the potential for filamentary synchrotron emission to emerge in relatively low-turbulence environments.

More generally in clusters, Low Frequency ARray (LOFAR: \citealt{van-Haarlem2013}) observations of Abell 2256 and the Coma Cluster have revealed filamentary structures in extended synchrotron emission on cluster scales, suggesting that such features may be more present than previously assumed \citep{van_Weeren_2019, Govoni_2019}. More recently, deep LOFAR observations of Abell 2199 have revealed multiple narrow, isolated synchrotron threads surrounding the radio lobes of 3C 338 \citep{Timmerman2026}. These isolated threads exhibit exceptionally steep spectral indices ($\alpha_{144}^{1500} \lesssim -3.0$) and have been interpreted as pre-existing magnetic flux tubes within the ICM that have subsequently captured runaway synchrotron-emitting plasma. These studies collectively highlight the importance of spectral index mapping in distinguishing between primary particle acceleration sites and regions dominated by synchrotron cooling \citep{Shimwell2017}. Filamentary features internal to lobes are well known, for example in the radio lobes of Cygnus A \citep{Perley1984}, a powerful Fanaroff-Riley type II (FRII) galaxy. These structures exhibit polarized emission, indicative of highly ordered magnetic fields. Similar findings have been reported for the Centaurus A radio galaxy, where filamentary structures are seen to align with the local magnetic field topology \citep{Anderson2018}. Such alignments suggest that magnetic reconnection or shearing processes may play a role in their formation, and indeed there is direct evidence to suggest that filaments within lobes are primarily magnetic enhancements rather than overdensities of synchrotron-emitting plasma \citep{Hardcastle2016}.

Filamentary structures may represent a transitional phase in the life cycle of radio galaxies, where ongoing particle acceleration and magnetic field interactions maintain synchrotron emission in aging lobes \citep{Ledlow1996}. Such features also provide evidence for episodic jet activity, as seen in sources such as Hercules A, where multiple generations of jets have left behind filamentary relics \citep{Gizani2003}. In the cluster context, filaments are indicative of the dynamic interplay between galaxies and their environments. The discovery of large-scale filaments in merging clusters, such as Abell 2255, suggests that cluster-wide magnetic fields play a role in shaping synchrotron structures \citep{Brunetti2014}. These findings also highlight the potential of filamentary structures as tracers of turbulence and magnetic field topology in the ICM.

ESO 137$-$006, observed with the MeerKAT array, revealed CSTs connecting its radio lobes, structures that were previously undetected in radio galaxies \citep{Ramatsoku2020}. The morphology and alignment of these CSTs suggest complex electromagnetic interactions that could be tied to the motion of the host galaxy through the ICM or reconnection events within magnetic field lines \citep{Heyvaerts1989}. However, the broader occurrence and underlying physical mechanisms of such features remain unexplored, especially beyond the scope of isolated, high-sensitivity studies.
Recent ASKAP observations of the nearby Corkscrew Galaxy in Abell 3627 reveal filamentary structures within its extended radio tail \citep{Koribalski2024}. These features, shaped by interactions with the surrounding ICM, exhibit similar properties to the CSTs in ESO 137$-$006, suggesting a possible relationship between these filaments and environmental effects such as turbulent gas motions, shocks, or magnetic field draping. The resemblance of these structures further strengthens the case for a common formation mechanism across different environments, reinforcing the need for high-resolution, multi-frequency studies to better understand their origins and evolution.

What is not yet known is the prevalence of these features in the general population of extended radio sources. To address this, the present paper leverages deep low-frequency, high-resolution observations from LOFAR. When combined with multi-frequency data, this enables a comprehensive investigation of this broader class of macroscopic filamentary structures, of which CSTs represent a highly collimated extreme. LOFAR's sensitivity to diffuse, steep-spectrum emission \citep{Shimwell2017} makes it well-suited for detecting faint and extended structures, while spectral analysis benefits from comparisons with higher-frequency radio data. This multi-wavelength approach allows for a detailed examination of the spectral properties of CSTs, which is crucial for distinguishing between different synchrotron aging processes and particle acceleration mechanisms \citep{Hardcastle2020}.

Investigating these narrow, elongated radio-emitting structures in a broader context holds the potential to answer important questions: Are these features indicative of a specific stage in the evolutionary cycle of radio galaxies? Do environmental factors, such as cluster density or galactic motion, play a significant role in their formation? Such studies are essential for advancing our understanding of jet-ICM interactions and the role of magnetic field dynamics within galaxy clusters \citep{Perucho2007, Mingo2014}.

For the purposes of this study, we define a general `synchrotron filament' as a spatially coherent, collimated feature that exhibits an enhanced surface brightness relative to the surrounding diffuse lobe emission or the intracluster medium (ICM). While our criteria encompass highly collimated threads analogous to those in ESO 137$-$006, we also target broader filamentary extensions that we identify as being structurally distinct from standard stochastic surface brightness fluctuations within the lobe plasma.

This paper presents the methodology and results of analysing such filamentary structures within a targeted sample of radio galaxies using LOFAR data, aiming to establish their occurrence rates, physical characteristics, and to put some constraints on their formation mechanisms. In the process, we report the discovery of new systems similar to ESO 137$-$006 and the Corkscrew galaxy and provide information on their large-scale radio and environmental properties.

To address distinct physical questions while maintaining structural clarity, this paper is separated into two independent parts. \textbf{Part I (Section \ref{sec:part1})} presents a broad morphological visual census aimed at constraining the statistical prevalence of synchrotron filaments in the general radio galaxy population. \textbf{Part II (Section \ref{sec:part2})} transitions to a purely physical analysis focusing exclusively on the three most robust, unambiguously resolved filament-hosting sources to derive minimum and thermal pressures. The paper concludes with a summary of both parts in Section \ref{sec:concl}.

\section{Part I: The Visual Census and Rarity of Synchrotron Filaments}
\label{sec:part1}

The initial objective of this study was to establish how common synchrotron filaments are across diverse cluster environments. The dataset for this study was derived from a combination of serendipitous discovery during visual inspections and systematic catalogue-based searches. This approach allowed for the identification and analysis of synchrotron filaments in diverse cluster environments.

\subsection{Source Identification and Initial Selection}
The investigation into these structures was initially motivated by the serendipitous discovery of two sources, ILTJ034927.68+751123.3 and ILTJ235728.23+475218.4, during visual inspections of the LoTSS Data Release 3 (LoTSS-DR3) imaging data \citep{Shimwell2026}. These sources exhibited distinct filamentary structures, highlighting the potential of visual inspection to uncover features that are not readily apparent in automated catalogue output source finders like PyBDSF \citep{Mohan2015}. Of these, ILTJ235728.23+475218.4 is very similar to ESO 137$-$006 in showing a classical wide-angle tail morphology but with a prominent filament that leaves and then re-enters the lobes, while ILTJ034927.68+751123.3 has a double-lobed morphology with filaments extending out of and back into as well as within the lobes.

To expand the sample systematically, public and proprietary datasets were employed, including:

\begin{enumerate}
    \item Data Release 2 (DR2): LoTSS data release (LoTSS-DR2); this dataset provided the foundation for identifications and redshift measurements for some sources, essential for deriving physical source properties. Radio data were taken from \cite{Shimwell2022} and optical identifications from \cite{Hardcastle2023}.
   \item AGN and Cluster Catalogs: We cross-matched the AGN catalogue of \cite{Hardcastle2025} \(\sim 600,000\) with a catalogue of environments of LoTSS-DR2 sources (Croston et al. submitted) which uses a galaxy-counting method to estimate the $M_{500}$ values of the environments of sources with a spectroscopic redshift in LoTSS-DR2. This yielded a refined subset of \(\sim 90,000\) potential candidates for filamentary structures which are simultaneously classified as AGN and have a host cluster/group mass estimate.
\end{enumerate}

\subsection{Selection Criteria}
To ensure the sample focused on sources with the highest potential for synchrotron filaments, a multi-stage filtering process was applied: we started from the AGN/cluster crossmatch catalogue of 90,000 sources, and then we applied the cuts below to systematically refine the sample.

\begin{enumerate}
    \item \textbf{Flux Density Thresholds:} Sources with flux densities exceeding 0.25 Jy at 144 MHz were prioritized, as lower flux densities often correspond to unresolved or diffuse sources unsuitable for detailed filament analysis. This threshold was selected to ensure sufficient signal-to-noise for resolving filamentary structures. Lower flux density sources were generally found to be too diffuse or unresolved for reliable filament detection. Empirical motivation for this cut comes from both previously studied and newly identified filament-hosting sources. For instance, ESO 137$-$006, observed with MeerKAT, exhibits peak brightness values of $\sim8$ mJy beam$^{-1}$ at 1030 MHz, where collimated synchrotron threads (CSTs) are clearly resolved \citep{Ramatsoku2020}. Similarly, the two serendipitously discovered LOFAR sources that served as initial filament candidates display peak brightnesses exceeding $6$ mJy beam$^{-1}$ at 144 MHz. Importantly, these sources are not only strong in surface brightness but also possess high total flux densities, reinforcing that both high surface brightness and high total flux are critical for reliably identifying and characterizing filamentary structures. These observations justify the adoption of a conservative $>0.5$ Jy at 144 MHz flux density threshold in our selection process to maintain consistency with known filament-hosting systems and to minimize false positives in lower-brightness, low-resolution environments.
    
    \item \textbf{Angular and Physical Size Constraints:} Candidates were required to have minimum angular sizes corresponding to physical extents of at least 500 kpc to distinguish extended radio structures from compact sources. This criterion was applied to maintain adequate resolution across each source, allowing filaments to be distinguished from diffuse lobes. Similarly, the choice of this threshold was guided by the characteristics of ESO 137$-$006 and in our two initial candidates identified from LoTSS DR3, where filamentary structures were found in extended radio galaxies rather than compact systems.
    
    \item \textbf{Cluster Environment Richness:} Since existing filaments have been found in objects in rich environments, we used the environmental mass from the catalogue of Croston et al. (submitted) to select for group and cluster environments with estimated $M_{500} > 10^{13} M_\odot$.
\end{enumerate}

The final DR2 cross-matched sample comprised 548 sources. We visually inspected the LOFAR images for all of these, and identified visible filamentary features in a total of 10 new sources (see the tiled visualizations in Fig.~\ref{fig:tiled_sources}). Combining our initial 2 serendipitous DR3 discoveries with the 10 candidates from our systematic DR2 search, our total identified sample comprises 12 sources (summarized in Table \ref{tab:lofar_sources_full}). However, of the 10 DR2 candidates, only one (ILTJ171106.89+394140.2) exhibited clearly resolved, isolated filaments suitable for detailed physical analysis alongside our two DR3 sources (shown in Fig.~\ref{fig:combined-sources}). This new target resembles the Corkscrew galaxy \citep{Koribalski2024} in that it is a tailed radio galaxy with strong evidence for interaction with its host environment. These three primary ``filament-hosting sources'' were selected for X-ray and minimum pressure analysis using \textit{ROSAT} data. The remaining 9 candidate sources displayed suggestive filamentary morphology but were either insufficiently resolved, too faint, or lacked the isolated structures necessary for a quantitative pressure comparison.

Redshift information for our new candidate sources was primarily obtained from the LOFAR DR2 catalogue, ensuring consistency in source identification and physical property estimation. However, redshifts for the two originally discovered sources were instead retrieved from the NASA Extragalactic Database (NED), as there is not yet an optical identification catalogue for DR3.

As part of our extended selection process, we examined sources across a range of cluster mass and flux density regimes to better understand the environmental and intrinsic conditions under which filamentary structures may arise. Our overall sample of 548 objects that were searched for filaments consisted of three subsamples. In our `main sample' -- consisting of 248 radio galaxies with flux densities above 0.5 Jy and residing in clusters with masses exceeding \(3 \times 10^{13}\, \mathrm{M}_\odot\) -- only six (\(\sim2.4\%\)) exhibited filament-like features upon visual inspection and were retained as candidate sources. In the `faint sample', comprising 193 sources with flux densities between 0.25 Jy and 0.5 Jy in similarly massive clusters, just three (\(\sim1.5\%\)) met our morphological criteria. For comparative purposes, we also evaluated a `poor sample' of 107 high-flux sources (flux $>0.5$ Jy) located in lower-mass clusters (\(1 \times 10^{13} < M_{500} < 3 \times 10^{13}\, \mathrm{M}_\odot\)), identifying only one (\(\sim0.9\%\)) with plausible filamentary emission. These statistics reinforce the rarity of synchrotron filaments even in the most favourable cluster and source conditions, suggesting that additional physical mechanisms -- beyond high mass or brightness alone -- play a critical role in filament formation. Although the numbers of sources are small, it is interesting that we can say at roughly the 90 per cent confidence level that the fraction of bright sources in rich environments showing filaments (6/248) is greater than the fraction in poor environments (1/107), using the Bayesian method described by \cite{Rickel+25}. This suggests that high-mass environments are more likely to host these structures. More data will be needed to draw more statistically significant conclusions.

\begin{figure*} 
    \centering
    \includegraphics[width=\textwidth, height=0.93\textheight, keepaspectratio]{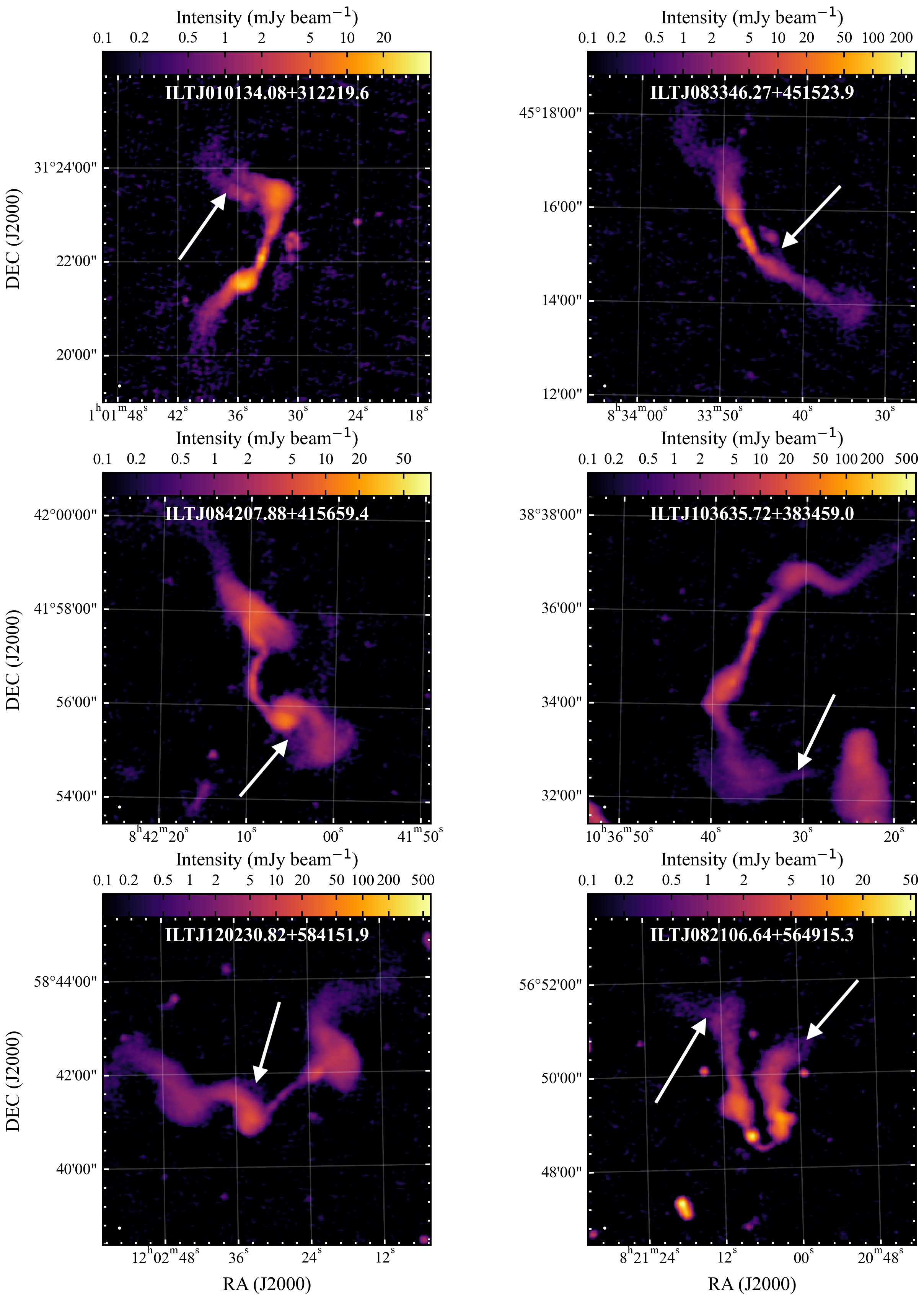}
    \caption{Tiled visualization of the ten candidate sources observed in the LOFAR Two-Metre Sky Survey (LoTSS). Each panel represents a distinct radio source, centered on its coordinates (RA, Dec) as derived from the survey catalog. The sources are plotted using a logarithmic color scale to enhance visibility of density variations. The images have been recentered to a fixed angular radius of 210 arcseconds, ensuring consistent spatial coverage. Overlaid grids indicate celestial coordinates. These visualizations provide a comparative view of the morphological structures of the sources, which are critical for analyzing their emission properties and potential classifications, e.g. as WAT radio galaxies, with the arrows pointing out possible filament positions. (Page 1 of 2)}
    \label{fig:tiled_sources}
\end{figure*}

\begin{figure*}[p]
    \addtocounter{figure}{-1} 
    \centering
    \includegraphics[width=\textwidth, height=0.63\textheight, keepaspectratio]{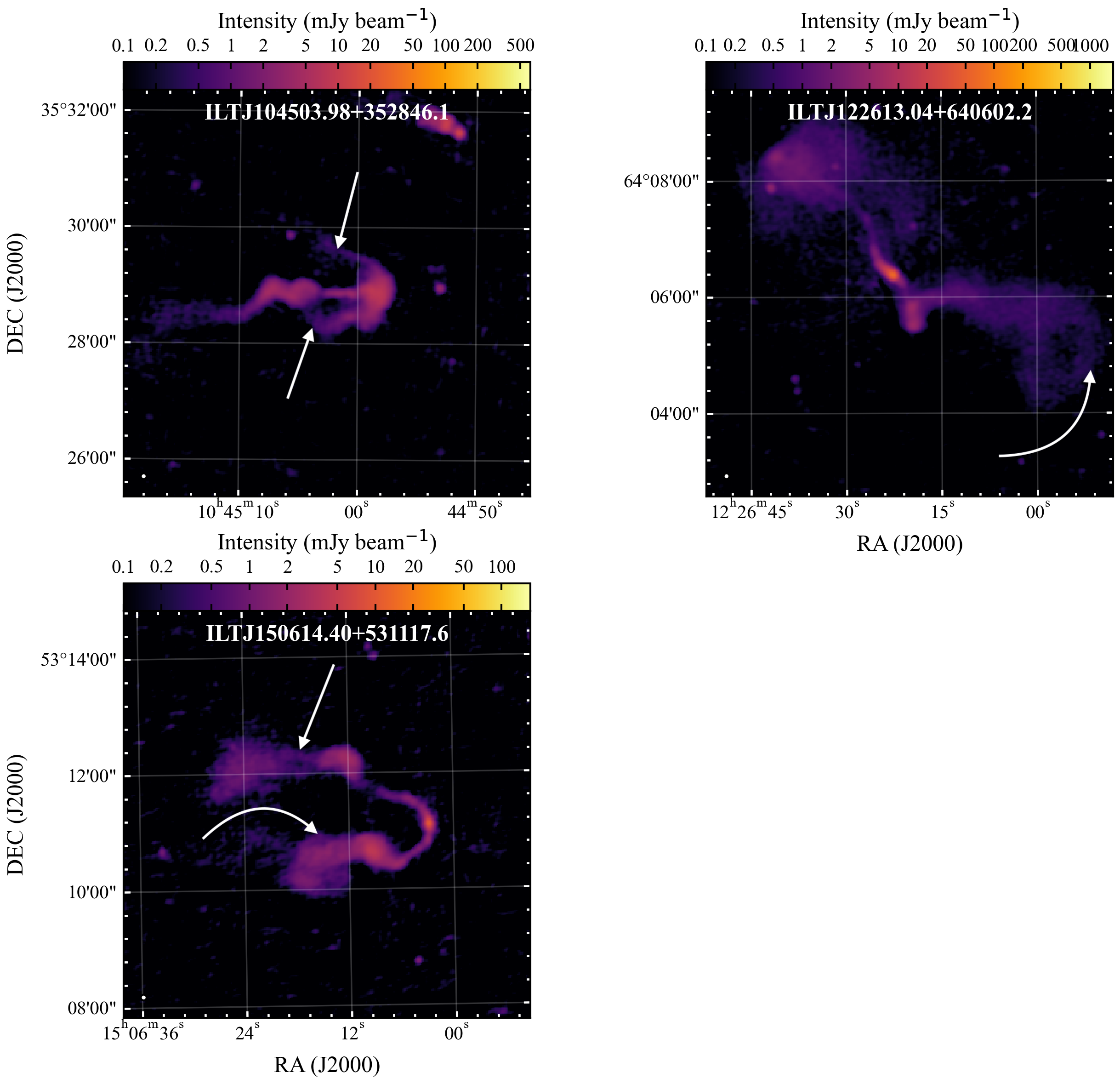}
    \caption{Tiled visualization of the ten candidate sources observed in the LOFAR Two-Metre Sky Survey (LoTSS). (continued).}
\end{figure*}

\begin{table*}
    \centering
    \caption{Summary of selected LOFAR sources and derived properties for the full sample.}
    \label{tab:lofar_sources_full}
    \begin{tabular}{lcccccccccc}
        \hline
        LOFAR Source Name & \(z\) & RA & DEC & \(\theta\) (arcsec) & \(D\) (kpc) & \(S_{150}\) (Jy) & \(S_{1.4}\) (Jy) & \(L_{150}\) (\(10^{25}\, \text{W/Hz}\)) & \(\alpha\) \\
        \hline
        ILTJ034927.68+751123.3 & 0.080 & 57.319 & 75.187 & 891 & 1332 & 6.150 & 0.045 & 11.45 & 2.16 \\
        ILTJ171106.89+394140.2 & 0.062 & 257.779 & 39.695 & 674 & 809 & 7.616 & 0.643 & 7.63 & 1.09 \\
        ILTJ235728.23+475218.4 & 0.045 & 359.368 & 47.872 & 698 & 645 & 4.255 & 0.585 & 2.17 & 0.87 \\
        ILTJ010134.08+312219.6 & 0.219 & 15.392 & 31.372 & 214 & 785 & 0.613 & 0.081 & 9.03 & 0.89 \\
        ILTJ083346.27+451523.9 & 0.180 & 128.443 & 45.257 & 283 & 643 & 0.589 & 0.083 & 5.58 & 0.86 \\
        ILTJ084207.88+415659.4 & 0.126 & 130.533 & 41.950 & 276 & 644 & 0.737 & 0.136 & 3.16 & 0.74 \\
        ILTJ103635.72+383459.0 & 0.145 & 159.149 & 38.583 & 530 & 1345 & 0.868 & 0.130 & 5.11 & 0.84 \\
        ILTJ120230.82+584151.9 & 0.127 & 180.628 & 58.698 & 500 & 1133 & 1.070 & 0.146 & 4.76 & 0.88 \\
        ILTJ082106.64+564915.3 & 0.173 & 125.278 & 56.821 & 284 & 530 & 0.506 & 0.067 & 4.42 & 0.89 \\
        ILTJ104503.98+352846.1 & 0.159 & 161.267 & 35.479 & 434 & 1192 & 0.320 & 0.037 & 2.35 & 0.95 \\
        ILTJ122613.04+640602.2 & 0.110 & 186.554 & 64.101 & 400 & 805 & 0.568 & 0.115 & 1.84 & 0.70 \\
        ILTJ150614.40+531117.6 & 0.145 & 226.559 & 53.188 & 298 & 785 & 0.302 & 0.042 & 1.79 & 0.87 \\
        \hline
    \end{tabular}
    \parbox{\textwidth}{\small \textbf{Note.} The first three entries correspond to the filament-hosting sources, while the remaining are candidate sources identified in this paper where filamentary features are less clearly resolved. Columns: \(z\) (redshift); RA and DEC (right ascension and declination, in degrees); \(\theta\) (angular size, in arcsec); \(D\) (physical size, in kpc); \(S_{150}\) and \(S_{1.4}\) are the LOFAR 144 MHz and NVSS 1.4 GHz integrated flux densities, respectively (in Jy); \(L_{150}\) is the 144 MHz radio luminosity (in \(10^{25}\,\text{W\,Hz}^{-1}\)); \(\alpha\) is the spectral index between 144 MHz and 1.4 GHz. Typical uncertainties: \(S_{150}\) \(\sim10\%\), \(S_{1.4}\) \(\sim3\%\); propagated uncertainty on \(\alpha\) is \(\sim0.1\); angular size uncertainty is \(\sim \pm 1\) arcsec due to 1.5 arcsec pixel resolution.}
\end{table*}

\subsection{Morphological Selection Criteria}

To classify synchrotron features as ‘filamentary’ for the purposes of this study, we employed a specific set of morphological criteria during the visual inspection of the LOFAR 144 MHz imagery. We defined a filament as a spatially coherent, collimated feature exhibiting a high length-to-width aspect ratio (typically $\gtrsim 3:1$), which appears distinct from the broader, amorphous emission characteristic of radio lobes. These structures typically manifest as narrow strands linking the main radio lobes or extending along the jet axis, rather than representing stochastic surface brightness fluctuations or local density enhancements within the wider lobe plasma \citep{Ramatsoku2020}.

Furthermore, we required that candidates demonstrate clear structural isolation to be selected for analysis. Valid filaments must be traceable as continuous features with enhanced surface brightness relative to the surrounding diffuse lobe emission or the intracluster medium (ICM), morphologically analogous to the ‘synchrotron threads’ observed in ESO 137$-$006 \citep{Ramatsoku2020}. This morphological filtering ensured that the final selection comprised only robust, physically distinct structures suitable for the independent aperture photometry and pressure analysis detailed in the subsequent sections.

We acknowledge that the identification of CSTs at the current LOFAR spatial resolution involves inherent morphological ambiguity. While certain candidates in our sample exhibit features that are unambiguously external to the main radio lobes, other filamentary characteristics may represent complex, internal magneto-hydrodynamic substructures. Distinguishing between true macroscopic CSTs traversing the intracluster medium and internal projection effects within the lobes remains a fundamental limitation of current low-frequency surveys. However, when studying a physical phenomenon, placing constraints on its prevalence is critical. Any systematic visual search yields astrophysical value by bounding the upper limits of occurrence. Even if some candidates in this sample suffer from projection effects, the search provides a definitive conclusion on the rarity of these structures in the broader population.

\subsection{Prevalence of Synchrotron Filaments}

Despite the morphological ambiguities inherent in individual candidates, the overarching statistical result of our uniform survey is robust: across a selected sample of 548 extended radio sources in rich environments, we find a complete absence of bright, unambiguous external filaments akin to those observed in systems like ESO 137$-$006. This scarcity is itself an important observational finding. It demonstrates that true macroscopic CSTs are exceptionally rare physical phenomena. This conclusion remains secure independent of whether our borderline candidates are ultimately classified as genuine external threads or internal lobe substructures due to projection effects, as the broader population clearly lacks widespread CST at least at the sensitivity of LOFAR.

The results reveal that synchrotron filaments are an exceptionally rare phenomenon within the observed population of radio galaxies. Among the extensive sample of approximately 548 active galactic nuclei (AGN) inspected, only a small fraction (\(\sim 2.2\%\)) exhibited candidate filamentary structures at a detectable level in the LOFAR observations. Filaments were more often found in sources that met strict flux density and cluster richness thresholds, suggesting that the combined effects of high environmental density and strong synchrotron emission are necessary but not sufficient conditions for filament formation. In environments where the estimated cluster mass was lower, there was a modestly significant decrease in the proportion of sources exhibiting filaments, underscoring the sensitivity of filament identification to both intrinsic source properties and external environmental conditions. This detection rate is likely influenced by the depth and resolution limits of the LOFAR imaging. Thin, diffuse filaments with low surface brightness may fall below the detection threshold or be resolved out, particularly in more distant or fainter systems. Similar observational challenges have been noted studies of non-thermal filaments \citep{Brienza2025}. A systematic comparison with sources without environmental density estimates, which may be in poorer environments, was not within the scope of this study but is planned for future work to further assess environmental dependencies in filament formation.

Interestingly, candidate filaments predominantly appeared in association with wide-angle tail galaxies, a morphology commonly linked to turbulent interactions between AGN jets and the intra-cluster medium (ICM) \citep{Hardcastle2020} and already linked to CSTs in systems like ESO 137$-$006 and 3C\,40B. Despite this apparent correlation, many well-studied wide-angle tail galaxies (e.g., 3C\,465; \citealt{Bempong_Manful_2020}) lack such structures, indicating that filament formation depends on additional, yet unidentified, mechanisms. These findings suggest that while filaments are not unique to any single class of sources, their occurrence is likely governed by complex environmental and dynamical factors. The rarity of these structures, combined with their dependence on both intrinsic and extrinsic conditions, highlights the need for further studies to uncover the physical processes that lead to their formation and persistence.

The results also indicate that filaments preferentially occur in rich group and cluster environments, where denser gas conditions and ICM interactions may facilitate their formation. The selection criteria were designed to prioritize sources with high flux densities ($>0.5$ Jy at 144 MHz) and extended morphologies ($>500$ kpc), allowing sufficient resolution to detect filamentary structures. However, the rarity of confirmed filaments, even in the most favorable environments, suggests that additional physical conditions --- such as local magnetic field configurations, shock compression, or past episodic AGN activity --- play a significant role in their formation and stability.

\section{Part II: Physical Analysis of Confirmed Filaments}
\label{sec:part2}

\begin{figure*}
    \centering
    \includegraphics[width=1\linewidth]{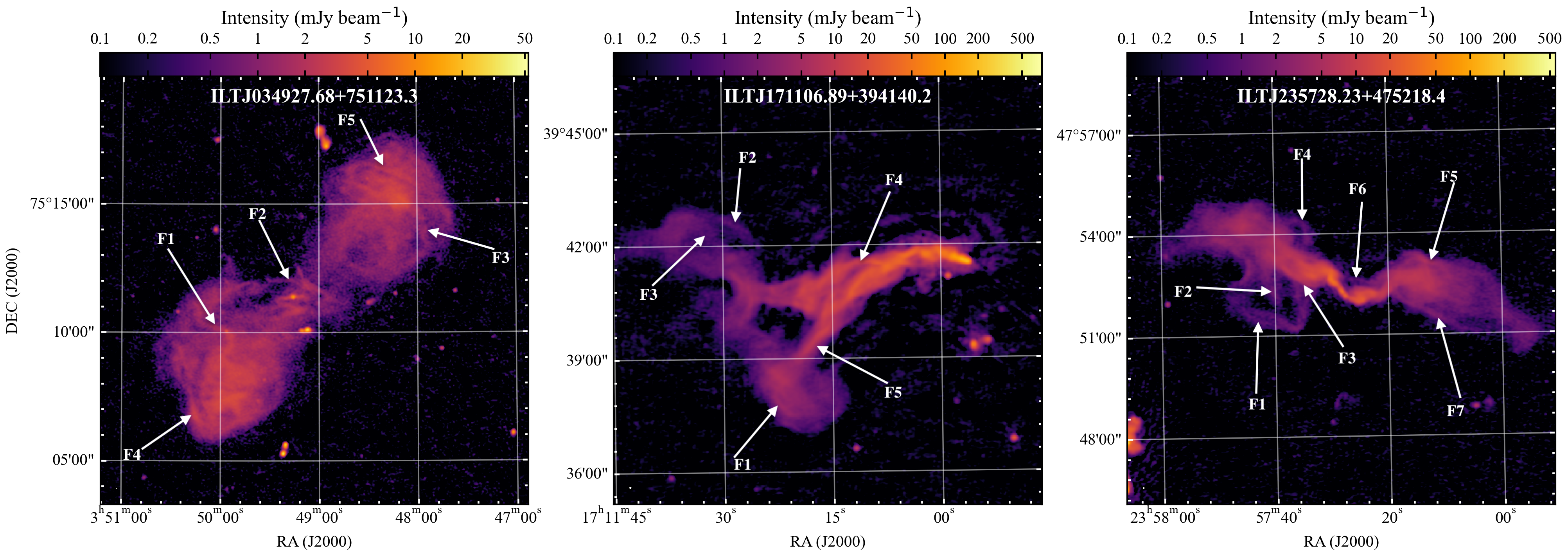}
    \caption{LoTSS view of the three primary filament-hosting sources: ILTJ034927.68+751123.3, ILTJ171106.89+394140.2, and ILTJ235728.23+475218.4. Maps are at 144 MHz with 6 arcsec resolution. The arrows provide a visual guide to the isolated filamentary structures.}
    \label{fig:combined-sources}
\end{figure*}

Having established the extreme scarcity of these features in the general population, the second part of this study transitions to a strictly independent physical and thermodynamic analysis. Here, we focus exclusively on the three most robust, high-confidence sources in our sample: ILTJ235728.23+475218.4, ILTJ171106.89+394140.2, and ILTJ034927.68+751123.3 (shown in Figure \ref{fig:combined-sources}). 

Because these specific targets possess clearly resolved, isolated filaments that are structurally uncoupled from the primary lobes, they are not subject to the morphological ambiguities discussed in Part I. The physical evaluation, X-ray environmental mapping, and minimum pressure models applied to these three systems stand entirely independent of the broader visual census. The methodology used in this study was to comprehensively characterize the physical and environmental properties of synchrotron filaments. This included a detailed analysis of flux densities, spectral indices, pressures, and magnetic field properties.

\subsection{Radio Flux Measurements and Spectral Index Calculation}
Radio flux densities were measured directly from LOFAR and NVSS survey images using aperture photometry techniques. This approach was necessary to account for spatial variations in extended structures, which are often misrepresented in catalog flux values. Flux densities at 144 MHz (LOFAR) and 1.4 GHz (NVSS) were used to calculate the spectral index ($\alpha$) for each source, allowing for a broad characterization of synchrotron aging trends across the identified filaments. Spectral indices provide insights into synchrotron aging and particle acceleration processes, with steeper spectra being indicative of older electron populations. We used the standard relation:

\begin{equation}
    \alpha = -\frac{\log(S_{\nu,1} / S_{\nu,2})}{\log(\nu_1 / \nu_2)}
\end{equation}

where $S_{\nu,1}$ and $S_{\nu,2}$ are the flux densities at LOFAR and NVSS frequencies respectively, to estimate the indices.

\subsection{X-ray Data Analysis}

The analysis of X-ray data, sourced from \textit{ROSAT} observations, was conducted to extract fluxes and calculate the thermal pressures of the intracluster medium (ICM). This section outlines the methodology used, including region definition, background subtraction, and flux conversion.

To extract source fluxes and constrain the thermal properties of the surrounding environments, regions were defined manually on LOFAR images using \texttt{SAOImage DS9}. Circular regions were placed to encompass the primary radio emission, with their sizes determined through visual inspection of the LOFAR source extent. Since the X-ray detections were weak and marginal in two cases, this targeted approach of importing the LOFAR-defined regions into the \textit{ROSAT} frames ensured consistent spatial alignment between the synchrotron-emitting regions and the corresponding X-ray extraction apertures.

For background estimation, separate source-free regions were selected in close proximity to the source regions. These background regions were chosen to reflect the local background levels, avoiding bright spots, unrelated sources, or detector artifacts. This ensured that the background contribution was representative of the observational environment near the source.

\begin{figure*}
    \centering
    \includegraphics[width=\textwidth]{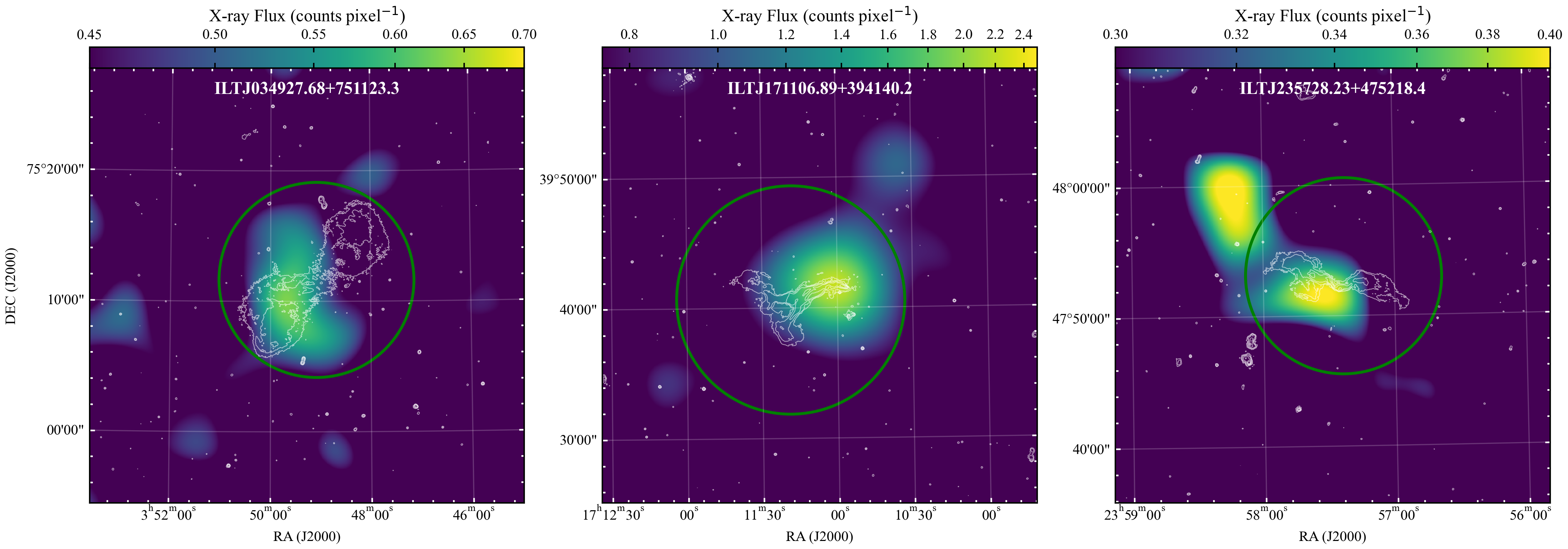}
    \caption{ROSAT} 0.1-2.4 keV X-ray maps of the three primary filament-hosting sources: ILTJ034927.68+751123.3 (left), ILTJ171106.89+394140.2 (center), and ILTJ235728.23+475218.4 (right). Overlaid are the LOFAR 144 MHz radio contours (white) to illustrate the radio source extent. The solid green regions indicate the apertures used for extracting the source X-ray counts.
    \label{fig:xray_maps}
\end{figure*}

Figure \ref{fig:xray_maps} presents the \textit{ROSAT} maps for our three primary targets, overlaid with LOFAR contours and our selected extraction apertures. This visually demonstrates the spatial relationship between the extended radio source morphologies, our photometric regions, and the broader cluster environment. The \textit{ROSAT} X-ray images were smoothed using a two-dimensional Gaussian kernel with a standard deviation of $\sigma = 80$ pixels.

Counts from the source regions were corrected for background contributions using the following equation:
\begin{equation}
    C_{\text{net}} = C_{\text{source}} - \frac{A_{\text{source}}}{A_{\text{background}}} \times C_{\text{background}},
\end{equation}
where $C_{\text{source}}$ and $C_{\text{background}}$ are the total counts in the source and background regions, respectively, and $A_{\text{source}}$ and $A_{\text{background}}$ are the areas of these regions in arcsec$^2$. This approach ensured an accurate estimation of the net source counts, eliminating potential biases due to background fluctuations.

The background-subtracted counts, $C_{\text{net}}$, were divided by the exposure time, $t_{\text{exp}}$, to calculate the count rate:
\begin{equation}
    R_{\text{count}} = \frac{C_{\text{net}}}{t_{\text{exp}}},
\end{equation}
where $t_{\text{exp}}$ is the exposure time in seconds. The count rate was subsequently converted into flux using the standard count-rate-to-flux conversion factor, $F_{\text{conv}}$, derived from \textit{ROSAT} calibration studies:
\begin{equation}
    F_X = R_{\text{count}} \cdot F_{\text{conv}},
\end{equation}
where $F_X$ represents the X-ray flux in units of erg~cm$^{-2}$~s$^{-1}$. For this study, we adopted a conversion factor $F_{\text{conv}} = 5.7 \times 10^{-12}$ erg~cm$^{-2}$~s$^{-1}$ per count, which assumes a thermal plasma model with a temperature of 3--6~keV and a metallicity of 0.3 solar \citep{Böhringer2014, Snowden1994}.

This strategy allowed us to extract constrained estimates of X-ray fluxes and luminosities under the assumption that any excess counts were associated with hot gas in the vicinity of the radio source. We emphasize that this method does not rely on radial profile fitting, but rather represents a luminosity-driven estimate of the ambient thermal environment. These count rates represent our best estimates of the hot gas emission from the radio source, whether or not they constitute significant detections. None of the sources discussed were found in standard \textit{ROSAT}-based cluster catalogs such as REFLEX or MCXC, reinforcing the idea that the derived fluxes represent local excesses associated with the radio galaxy environment, rather than known massive clusters. We note that potential contamination of the \textit{ROSAT} X-ray flux by unresolved nuclear active galactic nucleus (AGN) emission is expected to be minimal; standard radio-core to X-ray-core scaling relations for FR I radio galaxies (e.g., \citealt{Hardcastle1999}) indicate that expected nuclear X-ray counts are at least an order of magnitude below what is measured for each source when we use core flux density estimates from FIRST or VLASS. 

Statistical uncertainties in the counts were propagated through each step of the analysis to quantify the error in the final flux values (see Table~\ref{tab:xray_updated}). Uncertainties were derived from Poisson statistics for both the source and background regions, scaled to account for the relative areas of the extraction apertures. These were propagated through to the net count rates, fluxes, and luminosities to provide robust error estimates. For regions where the signal was weak or ambiguous, the background extraction and region boundaries were re-examined to ensure they were free from contamination or large-scale gradients, and the source apertures were refined to better encompass any diffuse emission. This procedure ensures that the derived X-ray quantities reflect a conservative and morphology-driven estimate of the thermal environment. Furthermore, the limited spatial and photon sensitivity of the \textit{ROSAT} All-Sky Survey introduces intrinsic scatter into our derived X-ray luminosities, which subsequently propagates as an additional systematic uncertainty into the estimated $M_{500}$ values.

\begin{table*}
\centering
\scriptsize
\caption{X-ray observational results for the synchrotron filament host galaxies, including propagated uncertainties. Count rates are derived from \textit{ROSAT} exposure times, and $M_{500}$ values are calculated using the luminosity scaling relation.}
\label{tab:xray_updated}
\begin{tabular}{lcccccc}
\hline
LOFAR Source Name & \textit{ROSAT} Obs. ID & Net Counts & Count Rates & Flux (erg\,cm$^{-2}$\,s$^{-1}$) & Luminosity (erg\,s$^{-1}$) & $M_{500}$ ($M_\odot$) \\
\hline
ILTJ235728.23+475218.4 & RS930843N00 & $21.30 \pm 10.36$ & $(3.55 \pm 1.73) \times 10^{-2}$ & $(2.02 \pm 0.98) \times 10^{-13}$ & $(1.04 \pm 0.51) \times 10^{42}$ & $(3.16 \pm 0.98) \times 10^{13}$ \\
ILTJ171106.89+394140.2 & RS931038N00 & $197.18 \pm 24.26$ & $(3.29 \pm 0.40) \times 10^{-1}$ & $(1.87 \pm 0.23) \times 10^{-12}$ & $(1.87 \pm 0.23) \times 10^{43}$ & $(2.01 \pm 0.16) \times 10^{14}$ \\
ILTJ034927.68+751123.3 & RS930404N00 & $33.61 \pm 14.76$ & $(5.17 \pm 2.27) \times 10^{-2}$ & $(2.95 \pm 1.29) \times 10^{-13}$ & $(5.01 \pm 2.20) \times 10^{42}$ & $(8.65 \pm 2.43) \times 10^{13}$ \\
\hline
\end{tabular}
\end{table*}

\subsection{\texorpdfstring{$M_{500}$ Estimation and Pressure Calculation from \textit{ROSAT}}{M500 Estimation and Pressure Calculation from \textit{ROSAT}}}

The parameter \( M_{500} \) represents the total mass enclosed within a radius \( R_{500} \), within which the mean density is 500 times the critical density of the Universe at the cluster redshift. It serves as a standard measure of cluster mass in X-ray and Sunyaev–Zel’dovich studies and is widely used to characterize the dynamical state and thermal content of galaxy clusters. In this study, \( M_{500} \) provides a basis for estimating the ambient thermal pressure of the intracluster medium (ICM) surrounding the radio sources, facilitating comparison with the synchrotron-derived minimum pressures of the embedded filaments.

The $M_{500}$ and $L_x$ relation follows work presented by \citet{Böhringer2014}. The scaling relation was established by correlating the X-ray luminosity ($L_x$) of galaxy clusters, measured in the \textit{ROSAT} $0.1\text{--}2.4 \, \text{keV}$ energy band, with their estimated masses ($M_{500}$) as in table \ref{tab:xray_updated}. The reference dataset consisted of clusters with known mass measurements, obtained via hydrostatic equilibrium models or weak lensing observations.

A power-law relation of the form $M_{500} = A \cdot \left( \frac{L_X}{L_0} \right)^{B}$ was adopted from \citet{Böhringer2014}, where $A$ is the normalization, $B$ is the slope, and $L_0 = 10^{42}$ erg s$^{-1}$ is the reference luminosity. Their analysis, which accounted for observational scatter and selection biases, yielded best-fit parameters of $A = 3.081 \times 10^{13} \, h_{70}^{-1} \, M_{\odot}$ and $B = 0.64$, providing a well-calibrated scaling relation for estimating $M_{500}$ from X-ray luminosity measurements.

\begin{equation}
      M_{500} = 3.081 \times 10^{13} \left( \frac{L_X}{10^{42} \, \mathrm{erg/s}} \right)^{0.64} \, h_{70}^{-1} \, M_{\odot}
\end{equation}

Previous work on individual sources uses radial profiling of the X-ray emission to study pressure and density profiles, but the existing X-ray data are not good enough to allow this. In this work, we instead employ \textit{ROSAT} X-ray data to derive $M_{500}$ estimates and use them to compute pressure profiles based on the universal pressure profile (UPP) \citep{Arnaud2010}. The pressure at a given radius $(R)$ was determined using the UPP model:

\begin{equation}
    P(R) = P_0 \cdot \left( \frac{c_{500} \cdot R}{R_{500}} \right)^{-\gamma} \cdot 
\left[ 1 + \left( \frac{c_{500} \cdot R}{R_{500}} \right)^{\alpha} \right]^{-(\beta - \gamma)/\alpha}
\end{equation}

where \( P_0 \) is the normalization constant, \( c_{500} \) is the concentration parameter, \( \gamma \) is the inner slope of the profile, \( \alpha \) controls the transition steepness between the inner and outer regions, \( \beta \) is the outer slope of the profile, and \( R_{500} \) is the radius within which the density is 500 times the critical density of the universe.

We adopt the best-fit values for these parameters from \citet{Arnaud2010}, obtained using \textit{XMM-Newton} observations of galaxy clusters. Although these parameters were originally calibrated on massive galaxy clusters, self-similar gravitational scaling ensures that they remain a valid and robust first-order approximation when applied to the lower-mass group and cluster environments encompassed in our sample \citep{Sun2011}.

\subsection{Synchrotron Modeling}
The synchrotron properties of the filaments were modeled using the Python-based {\sc pysynch} library \citep{Hardcastle-2018}, which calculates synchrotron emissivity under the assumption of power-law or aged electron distributions. The modeling parameters included the electron energy index, the energy range constraints, and the resulting magnetic field strengths and minimum pressures. Minimum pressures were estimated using the classical minimum energy condition assuming no protons, following the formalism of \citet{Hardcastle1998}, in which we integrate over a fixed energy range from $\gamma_{\rm min} = 1$ to $\gamma_{\rm max} = 10^{5}$ (cf. \citealt{Beck+Krause05}). The calculations adopted a fixed path length through the source based on the observed filament widths. These assumptions are consistent with prior work on FRI radio galaxies (e.g., \citealt{Croston2003}; \citealt{Hardcastle2005}) .

To assess the impact of the electron energy index, $p$, on synchrotron pressure estimates, we explored two representative cases—$p=2$ and $p=3$ corresponding to relatively flat and steep injection spectra, respectively. The electron population follows a power-law distribution, $N(E)\propto E^{-p}$ where the spectral index $\alpha$ relates to the particle index by $p=2\alpha+1$ \citep{Rybicki1979}. This comparison highlights the sensitivity of minimum-energy calculations to spectral assumptions, particularly in the absence of direct measurements for the filaments. As expected, adopting a steeper spectrum shifts more energy into low-energy electrons, resulting in significantly higher total energy densities and minimum pressures. For example, in filament J2357-f1, the pressure increases from $5.9\times10^{-15}$ Pa for $p=2$ to $1.4\times10^{-13}$ Pa for $p=3$, a factor of over 20. Across the sample, pressure ratios between the two cases typically range from 18 to 22. These results suggest that assuming a steeper spectral index, as might be expected due to radiative aging, may bring synchrotron-derived pressures more in line with those inferred from X-ray observations. Although direct spectral index measurements of the filaments in this study are not available, steep values are anticipated based on their morphology and by analogy with systems such as ESO 137$-$006, where similar features exhibit pronounced spectral steepening \citep{Ramatsoku2020}. The fact that the integrated spectra of our targets are steep (Table \ref{tab:lofar_sources_full}) supports the idea that the flat-spectrum minimum pressures may be underestimates. However, in what follows we retain the $p=2$ pressures as essentially providing a true minimum, noting that the pressure will be dominated by low-energy electrons that will not have undergone significant aging since their acceleration.

The derived minimum pressure ($P_{\mathrm{min}}$) is highly sensitive to several poorly constrained physical assumptions, particularly the precise emitting volume. Our baseline calculation assumes a uniform filling factor ($\phi = 1$) and, because we are explicitly estimating minimum pressures, no contribution from non-radiating particles. \cite{Hardcastle2013} showed that the presence of substantially non-uniform magnetic field strengths does not significantly affect the results of equipartition or minimum-energy calculations.

\subsection{Synchrotron Properties and Pressure Analysis}

The analysis indicates a consistent disparity between thermal and minimum pressures across the observed filaments, with thermal pressures being significantly higher (summarized in Table~\ref{tab:filaments_pressure}). This trend is observed in all systems studied, irrespective of their host environment or filament morphology. Filaments within the same system show varying pressures, suggesting localized environmental influences or intrinsic differences in their physical states. Such pressure imbalances have been reported in previous studies of radio galaxies, notably in sources like 3C\,449 and 3C\,465, where minimum pressures inferred from synchrotron emission were markedly lower than the surrounding ICM pressures derived from X-ray observations \citep{Hardcastle1998, Croston2003, Hardcastle2005}. These discrepancies are commonly attributed to the presence of non-radiating components such as relativistic protons or entrained thermal gas \citep{Croston2018}. The observed dominance of thermal pressure over minimum synchrotron pressure suggests that filaments are essentially in pressure balance with the local ICM, with thermal particles dominating the pressure both inside and outside the filament. This equilibrium may enable their persistence despite the dynamically active cluster environment. This, together with the similarity between the lobe and filament pressure properties discussed below, argues against any model in which the filaments are fundamentally different in nature or origin from the lobe material; it seems plausible that they are simply lobe material that in some way has become separated from the main lobe.

\begin{table}
    \centering
    \caption{minimum and Thermal Pressure for three filament-hosting sources and their lobes.}
    \label{tab:filaments_pressure}
    \begin{tabular}{lcccc} 
        \hline
        Filament/Lobe & \(R\) (kpc) & \(S_\nu\) (Jy) & \(P_{\text{eq}}\) (Pa) & \(P_{\text{th}}\) (Pa) \\
        \hline
        \multicolumn{5}{c}{ILTJ235728.23+475218.4} \\
        J2357-F1 & 148 & 0.324 & $1.43 \times 10^{-14}$ & $1.06 \times 10^{-13}$ \\
        J2357-F2 & 164 & 0.250 & $4.41 \times 10^{-14}$ & $9.08 \times 10^{-14}$ \\
        J2357-F3 & 120 & 0.209 & $2.81 \times 10^{-14}$ & $1.40 \times 10^{-13}$ \\
        J2357-F4 & 72  & 0.207 & $2.97 \times 10^{-14}$ & $2.79 \times 10^{-13}$ \\
        J2357-F5 & 124 & 0.204 & $4.62 \times 10^{-14}$ & $1.34 \times 10^{-13}$ \\
        J2357-F6 & 34  & 0.204 & $9.32 \times 10^{-14}$ & $4.63 \times 10^{-13}$ \\
        J2357-F7 & 161 & 0.192 & $2.57 \times 10^{-14}$ & $9.35 \times 10^{-14}$ \\
        J2357-L1 & 250 & 1.097 & $1.65 \times 10^{-14}$ & $4.53 \times 10^{-14}$ \\
        J2357-L2 & 240 & 1.338 & $1.31 \times 10^{-14}$ & $4.88 \times 10^{-14}$ \\
        \hline
        \multicolumn{5}{c}{ILTJ171106.89+394140.2} \\
        J1711-F1 & 305 & 0.074 & $1.21 \times 10^{-14}$ & $4.21 \times 10^{-13}$ \\
        J1711-F2 & 311 & 0.022 & $9.23 \times 10^{-15}$ & $4.07 \times 10^{-13}$ \\
        J1711-F3 & 520 & 0.024 & $7.89 \times 10^{-15}$ & $1.45 \times 10^{-13}$ \\
        J1711-F4 & 72  & 0.012 & $1.41 \times 10^{-14}$ & $2.15 \times 10^{-12}$ \\
        J1711-F5 & 232 & 0.010 & $8.34 \times 10^{-15}$ & $6.46 \times 10^{-13}$ \\
        J1711-L1 & 336 & 0.591 & $9.70 \times 10^{-15}$ & $3.55 \times 10^{-13}$ \\
        J1711-L2 & 381 & 0.244 & $7.22 \times 10^{-15}$ & $2.81 \times 10^{-13}$ \\
        \hline
        \multicolumn{5}{c}{ILTJ034927.68+751123.3} \\
        J0349-F1 & 277 & 0.215 & $1.36 \times 10^{-14}$ & $2.60 \times 10^{-13}$ \\
        J0349-F2 & 132 & 0.012 & $2.03 \times 10^{-14}$ & $2.34 \times 10^{-13}$ \\
        J0349-F3 & 513 & 0.076 & $1.70 \times 10^{-14}$ & $8.68 \times 10^{-14}$ \\
        J0349-F4 & 456 & 0.089 & $6.16 \times 10^{-14}$ & $9.85 \times 10^{-13}$ \\
        J0349-F5 & 130 & 0.041 & $4.60 \times 10^{-14}$ & $4.45 \times 10^{-13}$ \\
        J0349-L1 & 352 & 1.956 & $5.49 \times 10^{-15}$ & $9.70 \times 10^{-14}$ \\
        J0349-L2 & 622 & 1.661 & $5.05 \times 10^{-15}$ & $2.37 \times 10^{-14}$ \\
        \hline
    \end{tabular}
    \parbox{\columnwidth}{\small Note: \(R\) (Radius from center, kpc), \(S_\nu\) (Flux density, Jy), \(P_{\text{eq}}\) (Minimum pressure, Pa), \(P_{\text{th}}\) (Thermal pressure, Pa).}
\end{table}

\begin{figure*}
    \centering
    \includegraphics[width=1\linewidth]{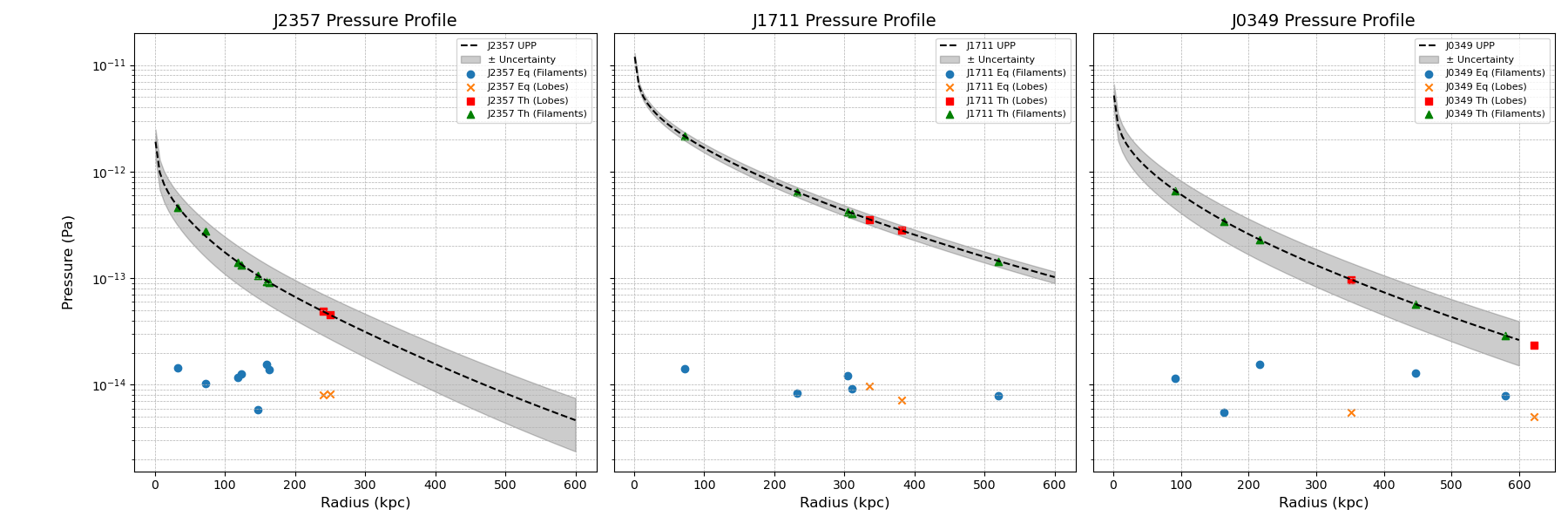}
    \caption{Universal Pressure Profile vs LOFAR-derived pressure estimates (thermal and minimum synchrotron) for our three filament-hosting sources. The Universal Pressure Profile \citep{Arnaud2010} is based on our estimates of the sources' environmental density from available X-ray data. The grey shaded region represents the uncertainty on pressures from the statistical uncertainty on the X-ray luminosity. The radio measurements come from the minimum pressure analysis of the LOFAR data. The very low pressures with respect to the external thermal pressure largely rule out an impulsive origin for the filaments and suggest that they originate as lobe material, while requiring a departure from equipartition, non-tangled magnetic fields and/or a pressure contribution from non-radiating particles to explain the filaments' continued existence.}
    \label{fig:pressure_profile}
\end{figure*}

\subsubsection{Radial Pressure Variations}

As illustrated by the universal pressure profiles in Fig.~\ref{fig:pressure_profile}, filaments located closer to the centers of their respective clusters exhibit systematically higher pressures compared to those situated at larger distances. This radial gradient reflects the influence of denser and hotter conditions in the cluster cores, which significantly affect the physical properties of the filaments. The trend holds across all observed systems, demonstrating the impact of the intra-cluster medium's central conditions on filament behavior. Additionally, because the observed filaments are not strictly aligned with the radial direction assumed by our spherical pressure profile model, projection effects and line-of-sight geometry introduce an extra source of uncertainty when comparing individual filament pressures against the radial thermal pressure profile.

The increasing pressure gradients toward the cluster cores reflect the denser and hotter conditions in the central regions of the ICM. These conditions likely enhance the confinement and stability of synchrotron filaments. Previous studies, such as those by \citet{van_Weeren_2019}, also report higher synchrotron emissivity and aging in core regions, consistent with these observations. The elevated pressures may also arise from shocks or dynamic interactions between cluster mergers, as suggested by \citet{Brunetti2014}, where magnetic reconnection or turbulent amplification contributes to filament stabilization in these high-pressure environments.

\subsubsection{Comparison Between Filaments and Lobes}

The lobes of these systems consistently exhibit minimum pressures that are comparable to, but somewhat lower than, their associated filaments, suggesting differences in their dynamical and magnetic environments, though the exact nature of these differences remains uncertain. While lobes represent more diffuse and less confined regions, filaments may not necessarily be sites of higher overall pressure, but rather regions of enhanced magnetic field strength, leading to increased synchrotron emissivity without a significant increase in total energy density. If the filaments were genuinely over-pressured relative to the lobes, material outflows into the surrounding lobe plasma would be expected, for which there is no observational evidence. Instead, the apparent pressure difference could arise from localized magnetic field amplification due to compression, shear flows, or interactions with the surrounding intracluster medium (ICM).

Toroidal magnetic field configurations, as discussed by \citet{Perley1984}, could provide additional rigidity, helping filaments maintain their structure despite external perturbations. However, their exact origin and evolutionary pathway remain open questions. One possibility is that filaments represent remnant structures of past AGN activity, where jet material has been funneled into highly magnetized strands through a combination of relic plasma interactions, weak shocks, or turbulence within the ICM. This shock-driven scenario is quantitatively supported by recent models of the CSTs in ESO 137$-$006 by \citet{Alam2025}, who demonstrated that the dynamical timescale associated with the local shock front ($\sim 70$ Myr) is similar to the estimated synchrotron age of the threads ($\sim 130$ Myr). The fact that we see that the filaments are very similar to local lobe material in their properties may support this type of model. Alternatively, filaments may form due to differential entrainment of magnetized plasma from the surrounding environment, selectively enhancing synchrotron emissivity in certain regions while leaving adjacent lobe material less affected. An alternative perspective, recently proposed by \citet{GopalKrishna2024}, suggests that filaments may arise from plasma instabilities and localized charge-separation effects within the radio lobes, rather than purely as relics of previous jet activity. In this scenario, large-scale electric fields and dynamic magneto-ionic interactions could drive the formation of synchrotron-emitting filaments, implying that they are not merely passive structures but instead result from ongoing plasma processes. However, if such a mechanism were dominant, it is unclear why the observed filament pressures appear broadly comparable to those of the surrounding lobes --- there is no a priori reason to expect local plasma processes to yield energy densities finely tuned to the broader lobe environment. Understanding whether filaments serve as conduits for energy transport between jets and lobes or instead arise as isolated, self-regulated structures will require detailed constraints on their kinematics and magnetic configurations, particularly from multi-frequency spectral imaging and Faraday rotation studies.

\section{Conclusion}
\label{sec:concl}

This study provides a comprehensive analysis of synchrotron filaments and their interplay with the intracluster medium (ICM), using multi-frequency radio and X-ray observations. In Part I, we carried out a systematic search for filaments in bright radio sources in rich environments in LoTSS DR2: our key conclusion is that filamentary structures like those recently revealed in ESO 137$-$006 are rare, with only 10 candidate filamentary sources in total out of 548 that we visually inspected. Many, but not all of our candidates are wide-angle tailed sources like ESO 137$-$006 and 3C\,40B, and they show a weak statistical preference for the richer environments in the cluster catalog.

In Part II we then investigated the nature of the filaments using the three newly found LOFAR sources with the best evidence for filamentary structure. Using archival \textit{ROSAT} observations, we made an estimate of the cluster luminosity and hence mass and used it to show that the filaments, like the lobes, of these objects have minimum pressures that lie substantially below the plausible level of the thermal pressure in their environments.

The comparison of synchrotron and thermal pressures highlights the quasi-static equilibrium of filaments within the ICM, with synchrotron pressures systematically lower than their thermal counterparts. This equilibrium and the slightly higher minimum pressures in the filaments with resepect to the lobes supports the hypothesis that these filaments are pre-existing lobe material that is stabilized within the cluster environment by magnetic tension forces, confining relativistic particles while resisting external perturbations. Moreover, our pressure profile analysis underscores the importance of environmental factors in shaping the morphology and dynamics of these structures. Models in which the filaments are fundamentally different in nature or origin from the lobes are disfavored because we would not expect to see the observed similarities in the behavior of the filaments and the lobes.

Looking forward, future investigations will benefit from multi-frequency polarimetric observations that will allow us to measure spectral index and polarization properties of the filaments, combined with and advanced numerical simulations to refine our understanding of magnetic field amplification and synchrotron filament evolution. Recent 3D general relativistic magnetohydrodynamic (GRMHD) simulations have begun to bridge the gap between observation and theory, demonstrating that CST-like features can emerge dynamically via shear-driven Kelvin-Helmholtz instabilities along cocoon boundaries or through magnetized back-flows, where retreating plasma drags magnetic flux downstream into the ICM \citep{Fromm2026}. Next-generation facilities such as the Square Kilometer Array (SKA) will provide insights into the complex interplay between AGN feedback, magnetic fields, and the ICM, allowing us to test these competing physical models against observational data in large samples. These efforts will further our understanding of the role of synchrotron filaments as tracers of cluster dynamics, particle acceleration, and the large-scale structure of the universe.

\section*{Acknowledgements}

GO acknowledges financial support from Development in Africa with Radio Astronomy (DARA) through a PhD studentship [ST/Y006100/1]. MJH thanks the UK STFC for support [ST/Y001249/1]. We are grateful to Maya Horton for comments on the manuscript.

LOFAR is the Low Frequency Array, designed and constructed by ASTRON. It has observing, data processing, and data storage facilities in several countries, which are owned by various parties (each with their own funding sources), and which are collectively operated by the ILT foundation under a joint scientific policy. The ILT resources have benefited from the following recent major funding sources: CNRS-INSU, Observatoire de Paris and Université d'Orléans, France; BMBF, MIWF-NRW, MPG, Germany; Science Foundation Ireland (SFI), Department of Business, Enterprise and Innovation (DBEI), Ireland; NWO, The Netherlands; The Science and Technology Facilities Council, UK; Ministry of Science and Higher Education, Poland; The Istituto Nazionale di Astrofisica (INAF), Italy.

This research made use of the Dutch national e-infrastructure with support of the SURF Cooperative (e-infra 180169) and the LOFAR e-infra group. The Jülich LOFAR Long Term Archive and the German LOFAR network are both coordinated and operated by the Jülich Supercomputing Centre (JSC), and computing resources on the supercomputer JUWELS at JSC were provided by the Gauss Centre for Supercomputing e.V. (grant CHTB00) through the John von Neumann Institute for Computing (NIC).

This research made use of the University of Hertfordshire high-performance computing facility and the LOFAR-UK computing facility located at the University of Hertfordshire (\url{https://uhhpc.herts.ac.uk}) and supported by STFC [ST/P000096/1], and of the Italian LOFAR IT computing infrastructure supported and operated by INAF, and by the Physics Department of Turin University (under an agreement with Consorzio Interuniversitario per la Fisica Spaziale) at the C3S Supercomputing Centre, Italy.

\section*{Data Availability}

The data used in this work are part of a LoTSS DR2 sample which can be accessed from \url{http://lofar-surveys.org}, \textit{ROSAT} and NVSS archival data accessed from \url{https://heasarc.gsfc.nasa.gov}.

\bibliographystyle{mnras}
\bibliography{ref}

@ARTICLE{Hardcastle2013,
       author = {{Hardcastle}, M.~J.},
        title = "{Synchrotron and inverse-Compton emission from radio galaxies with non-uniform magnetic field and electron distributions}",
      journal = {\mnras},
         year = 2013,
        month = aug,
       volume = {433},
       number = {4},
        pages = {3364-3372},
          doi = {10.1093/mnras/stt1024},
archivePrefix = {arXiv},
       eprint = {1306.1640},
 primaryClass = {astro-ph.HE},
       adsurl = {https://ui.adsabs.harvard.edu/abs/2013MNRAS.433.3364H}
}

@ARTICLE{Croston2018,
       author = {{Croston}, J.~H. and {Ineson}, J. and {Hardcastle}, M.~J.},
        title = "{Particle content, radio-galaxy morphology, and jet power: all radio-loud AGN are not equal}",
      journal = {\mnras},
         year = 2018,
        month = may,
       volume = {476},
       number = {2},
        pages = {1614-1623},
          doi = {10.1093/mnras/sty274},
archivePrefix = {arXiv},
       eprint = {1801.10172},
 primaryClass = {astro-ph.GA},
       adsurl = {https://ui.adsabs.harvard.edu/abs/2018MNRAS.476.1614C}
}

@ARTICLE{Hardcastle2023,
       author = {{Hardcastle}, M.~J. and {Horton}, M.~A. and {Williams}, W.~L. and {Duncan}, K.~J. and {Alegre}, L. and {Barkus}, B. and {Croston}, J.~H. and {Dickinson}, H. and {Osinga}, E. and {R{\"o}ttgering}, H.~J.~A. and {Sabater}, J. and {Shimwell}, T.~W. and {Smith}, D.~J.~B. and {Best}, P.~N. and {Botteon}, A. and {Br{\"u}ggen}, M. and {Drabent}, A. and {de Gasperin}, F. and {G{\"u}rkan}, G. and {Hajduk}, M. and {Hale}, C.~L. and {Hoeft}, M. and {Jamrozy}, M. and {Kunert-Bajraszewska}, M. and {Kondapally}, R. and {Magliocchetti}, M. and {Mahatma}, V.~H. and {Mostert}, R.~I.~J. and {O'Sullivan}, S.~P. and {Pajdosz-{\'S}mierciak}, U. and {Petley}, J. and {Pierce}, J.~C.~S. and {Prandoni}, I. and {Schwarz}, D.~J. and {Shulewski}, A. and {Siewert}, T.~M. and {Stott}, J.~P. and {Tang}, H. and {Vaccari}, M. and {Zheng}, X. and {Bailey}, T. and {Desbled}, S. and {Goyal}, A. and {Gonano}, V. and {Hanset}, M. and {Kurtz}, W. and {Lim}, S.~M. and {Mielle}, L. and {Molloy}, C.~S. and {Roth}, R. and {Terentev}, I.~A. and {Torres}, M.},
        title = "{The LOFAR Two-Metre Sky Survey. VI. Optical identifications for the second data release}",
      journal = {\aap},
         year = 2023,
        month = oct,
       volume = {678},
          eid = {A151},
        pages = {A151},
          doi = {10.1051/0004-6361/202347333},
archivePrefix = {arXiv},
       eprint = {2309.00102},
 primaryClass = {astro-ph.GA},
       adsurl = {https://ui.adsabs.harvard.edu/abs/2023A&A...678A.151H}
}

@ARTICLE{Shimwell2026,
       author = {{Shimwell}, T.~W. and {Hardcastle}, M.~J. and {Tasse}, C. and {Drabent}, A. and {Botteon}, A. and {Williams}, W.~L. and {Best}, P.~N. and {R{\"o}ttgering}, H.~J.~A. and {Br{\"u}ggen}, M. and {Brunetti}, G. and {Callingham}, J.~R. and {Chy{\.z}y}, K.~T. and {Conway}, J.~E. and {De Gasperin}, F. and {Haverkorn}, M. and {Horellou}, C. and {Jackson}, N. and {Miley}, G.~K. and {Morabito}, L.~K. and {Morganti}, R. and {O'Sullivan}, S.~P. and {Schwarz}, D.~J. and {Smith}, D.~J.~B. and {van Weeren}, R.~J. and {Vedantham}, H.~K. and {White}, G.~J. and {Ahmadi}, A. and {Alegre}, L. and {Arias}, M. and {Asabere}, B. and {Bahr-Kalus}, B. and {Barkus}, B. and {Bilicki}, M. and {B{\"o}hme}, L. and {Brentjens}, M. and {Brienza}, M. and {Bomans}, D.~J. and {Bonafede}, A. and {Bonato}, M. and {Bonnassieux}, E. and {Boxelaar}, J.~M. and {Camera}, S. and {Cassano}, R. and {Chilufya}, J. and {Cianfaglione}, M. and {Croston}, J.~H. and {Cuciti}, V. and {Dabhade}, P. and {De Rubeis}, E. and {de Jong}, J.~M.~G.~H.~J. and {Dallacasa}, D. and {Dettmar}, R.~J. and {Duncan}, K.~J. and {Di Gennaro}, G. and {Edler}, H.~W. and {Groeneveld}, C. and {G{\"u}rkan}, G. and {Hajduk}, M. and {Hale}, C.~L. and {Heesen}, V. and {Hoang}, D.~N. and {Hoeft}, M. and {Holties}, H. and {Horton}, M.~A. and {Iacobelli}, M. and {Jamrozy}, M. and {Jarvis}, M.~J. and {Jelic}, V. and {Kadler}, M. and {Kondapally}, R. and {Kunert-Bajraszewska}, M. and {Loose}, M. and {Magliocchetti}, M. and {Ma{\l}ek}, K. and {Manzano}, C. and {McKean}, J.~P. and {Mevius}, M. and {Mingo}, B. and {Miskolczi}, A. and {Misra}, A. and {Mold{\'o}n}, J. and {Nair}, D.~G. and {Nakoneczny}, S.~J. and {Orru}, E. and {Pashapour-Ahmadabadi}, M. and {Pasini}, T. and {Petley}, J. and {Pierce}, J.~C.~S. and {Prandoni}, I. and {Rafferty}, D. and {Rajpurohit}, K. and {Riseley}, C.~J. and {Roberts}, I.~D. and {Sethi}, S. and {Shulevski}, A. and {Stein}, M. and {Stuardi}, C. and {Sweijen}, F. and {ter Veen}, S. and {Timmerman}, R. and {Vaccari}, M. and {Wijnholds}, S.},
        title = "{The LOFAR Two-metre Sky Survey: VII. Third Data Release}",
      journal = {\aap},
         year = 2026,
        month = mar,
       volume = {707},
          eid = {A198},
        pages = {A198},
          doi = {10.1051/0004-6361/202557749},
archivePrefix = {arXiv},
       eprint = {2602.15949},
 primaryClass = {astro-ph.GA},
       adsurl = {https://ui.adsabs.harvard.edu/abs/2026A&A...707A.198S}
}

@ARTICLE{Sun2011,
       author = {{Sun}, M. and {Sehgal}, N. and {Voit}, G.~M. and {Donahue}, M. and {Jones}, C. and {Forman}, W. and {Vikhlinin}, A. and {Sarazin}, C.},
        title = "{The Pressure Profiles of Hot Gas in Local Galaxy Groups}",
      journal = {\apjl},
         year = 2011,
        month = feb,
       volume = {727},
       number = {2},
          eid = {L49},
        pages = {L49},
          doi = {10.1088/2041-8205/727/2/L49},
archivePrefix = {arXiv},
       eprint = {1012.0312},
 primaryClass = {astro-ph.CO},
       adsurl = {https://ui.adsabs.harvard.edu/abs/2011ApJ...727L..49S}
}

@ARTICLE{Fromm2026,
       author = {{Fromm}, Christian M. and {Kadler}, Matthias and {Mannheim}, Karl},
        title = "{Exploring the physics behind the observed magnetic filaments in large scale radio galaxies}",
      journal = {arXiv e-prints},
         year = 2026,
        month = jun,
          eid = {arXiv:2606.27591},
        pages = {arXiv:2606.27591},
          doi = {10.48550/arXiv.2606.27591},
archivePrefix = {arXiv},
       eprint = {2606.27591},
 primaryClass = {astro-ph.HE},
       adsurl = {https://ui.adsabs.harvard.edu/abs/2026arXiv260627591F}
}

@ARTICLE{Alam2025,
       author = {{Alam}, Toushif and {Mooley}, Kunal P. and {Sarkar}, Kartick C.},
        title = "{Origin of the {\ensuremath{\sim}}150 kpc radio filament in galaxy ESO 137{\ensuremath{-}}006}",
      journal = {\mnras},
         year = 2025,
        month = sep,
       volume = {542},
       number = {2},
        pages = {1465-1476},
          doi = {10.1093/mnras/staf1266},
archivePrefix = {arXiv},
       eprint = {2507.23026},
 primaryClass = {astro-ph.GA},
       adsurl = {https://ui.adsabs.harvard.edu/abs/2025MNRAS.542.1465A}
}

@ARTICLE{Timmerman2026,
       author = {{Timmerman}, R. and {Rudnick}, L. and {Botteon}, A. and {Brunetti}, G. and {Kale}, R.},
        title = "{The magnetic mayhem in Abell 2199: discovery of synchrotron threads and homogeneous diffuse radio lobes}",
      journal = {\mnras},
         year = 2026,
        month = aug,
       volume = {550},
       number = {2},
          eid = {stag1232},
        pages = {stag1232},
          doi = {10.1093/mnras/stag1232},
archivePrefix = {arXiv},
       eprint = {2606.30714},
 primaryClass = {astro-ph.CO},
       adsurl = {https://ui.adsabs.harvard.edu/abs/2026MNRAS.550g1232T}
}

@ARTICLE{Hardcastle1999,
       author = {{Hardcastle}, M.~J. and {Worrall}, D.~M.},
        title = "{ROSAT X-ray observations of 3CRR radio sources}",
      journal = {\mnras},
         year = 1999,
        month = nov,
       volume = {309},
       number = {4},
        pages = {969-990},
          doi = {10.1046/j.1365-8711.1999.02945.x},
archivePrefix = {arXiv},
       eprint = {astro-ph/9907034},
 primaryClass = {astro-ph},
       adsurl = {https://ui.adsabs.harvard.edu/abs/1999MNRAS.309..969H}
}

@ARTICLE{Rickel+25,
       author = {{Rickel}, Mary and {Moravec}, Emily and {Gordon}, Yjan A. and {Hardcastle}, Martin J. and {Pierce}, Jonathon C.~S. and {Bilton}, Lawrence E. and {Roberts}, Ian D.},
        title = "{The Merging Galaxy Cluster Environment Affects the Morphology of Radio Active Galactic Nuclei}",
      journal = {\apj},
         year = 2025,
        month = apr,
       volume = {983},
       number = {2},
          eid = {138},
        pages = {138},
          doi = {10.3847/1538-4357/adbb5e},
archivePrefix = {arXiv},
       eprint = {2502.04198},
 primaryClass = {astro-ph.GA},
       adsurl = {https://ui.adsabs.harvard.edu/abs/2025ApJ...983..138R}
}

@ARTICLE{Beck+Krause05,
       author = {{Beck}, R. and {Krause}, M.},
        title = "{Revised equipartition and minimum energy formula for magnetic field strength estimates from radio synchrotron observations}",
      journal = {Astronomische Nachrichten},
         year = 2005,
        month = jul,
       volume = {326},
       number = {6},
        pages = {414-427},
          doi = {10.1002/asna.200510366},
archivePrefix = {arXiv},
       eprint = {astro-ph/0507367},
 primaryClass = {astro-ph},
       adsurl = {https://ui.adsabs.harvard.edu/abs/2005AN....326..414B}
}

@ARTICLE{Rudnick2022,
       author = {{Rudnick}, L. and {Br{\"u}ggen}, M. and {Brunetti}, G. and {Cotton}, W.~D. and {Forman}, W. and {Jones}, T.~W. and {Nolting}, C. and {Schellenberger}, G. and {van Weeren}, R.},
        title = "{Intracluster Magnetic Filaments and an Encounter with a Radio Jet}",
      journal = {\apj},
         year = 2022,
        month = aug,
       volume = {935},
       number = {2},
          eid = {168},
        pages = {168},
          doi = {10.3847/1538-4357/ac7c76},
archivePrefix = {arXiv},
       eprint = {2206.14319},
 primaryClass = {astro-ph.HE},
       adsurl = {https://ui.adsabs.harvard.edu/abs/2022ApJ...935..168R}
}

@ARTICLE{Sakelliou2008,
       author = {{Sakelliou}, Irini and {Hardcastle}, M.~J. and {Jetha}, N.~N.},
        title = "{3C40 in Abell194: can tail radio galaxies exist in a quiescent cluster?}",
      journal = {\mnras},
         year = 2008,
        month = feb,
       volume = {384},
       number = {1},
        pages = {87-93},
          doi = {10.1111/j.1365-2966.2007.12465.x},
archivePrefix = {arXiv},
       eprint = {0709.2133},
 primaryClass = {astro-ph},
       adsurl = {https://ui.adsabs.harvard.edu/abs/2008MNRAS.384...87S}
}

@ARTICLE{Brienza2025,
       author = {{Brienza}, M. and {Rajpurohit}, K. and {Churazov}, E. and {Heywood}, I. and {Br{\"u}ggen}, M. and {Hoeft}, M. and {Vazza}, F. and {Bonafede}, A. and {Botteon}, A. and {Brunetti}, G. and {Gastaldello}, F. and {Khabibullin}, I. and {Lyskova}, N. and {Majumder}, A. and {R{\"o}ttgering}, H.~J.~A. and {Shimwell}, T.~W. and {Simionescu}, A. and {van Weeren}, R.~J.},
        title = "{Non-thermal filaments and AGN recurrent activity in the galaxy group Nest200047: A LOFAR, uGMRT, MeerKAT, and VLA radio spectral analysis}",
      journal = {\aap},
         year = 2025,
        month = apr,
       volume = {696},
          eid = {A239},
        pages = {A239},
          doi = {10.1051/0004-6361/202553676},
archivePrefix = {arXiv},
       eprint = {2502.18244},
 primaryClass = {astro-ph.GA},
       adsurl = {https://ui.adsabs.harvard.edu/abs/2025A&A...696A.239B}
}

@ARTICLE{Hardcastle2005,
       author = {{Hardcastle}, M.~J. and {Sakelliou}, I. and {Worrall}, D.~M.},
        title = "{A Chandra and XMM-Newton study of the wide-angle tail radio galaxy 3C465}",
      journal = {\mnras},
         year = 2005,
        month = may,
       volume = {359},
       number = {3},
        pages = {1007-1021},
          doi = {10.1111/j.1365-2966.2005.08966.x},
archivePrefix = {arXiv},
       eprint = {astro-ph/0502575},
 primaryClass = {astro-ph},
       adsurl = {https://ui.adsabs.harvard.edu/abs/2005MNRAS.359.1007H}
}

@ARTICLE{Croston2003,
       author = {{Croston}, J.~H. and {Hardcastle}, M.~J. and {Birkinshaw}, M. and {Worrall}, D.~M.},
        title = "{XMM-Newton observations of the hot-gas atmospheres of 3C 66B and 3C 449}",
      journal = {\mnras},
         year = 2003,
        month = dec,
       volume = {346},
       number = {4},
        pages = {1041-1054},
          doi = {10.1111/j.1365-2966.2003.07165.x},
archivePrefix = {arXiv},
       eprint = {astro-ph/0309150},
 primaryClass = {astro-ph},
       adsurl = {https://ui.adsabs.harvard.edu/abs/2003MNRAS.346.1041C}
}

@ARTICLE{Hardcastle1998,
       author = {{Hardcastle}, M.~J. and {Worrall}, D.~M. and {Birkinshaw}, M.},
        title = "{Dynamics of the radio galaxy 3C449}",
      journal = {\mnras},
         year = 1998,
        month = jun,
       volume = {296},
       number = {4},
        pages = {1098-1104},
          doi = {10.1046/j.1365-8711.1998.01535.x},
archivePrefix = {arXiv},
       eprint = {astro-ph/9802175},
 primaryClass = {astro-ph},
       adsurl = {https://ui.adsabs.harvard.edu/abs/1998MNRAS.296.1098H}
}

@misc{Mohan2015,
       author = {{Mohan}, Niruj and {Rafferty}, David},
        title = "{PyBDSF: Python Blob Detection and Source Finder}",
 howpublished = {Astrophysics Source Code Library, record ascl:1502.007},
         year = 2015,
        month = feb,
          eid = {ascl:1502.007},
       adsurl = {https://ui.adsabs.harvard.edu/abs/2015ascl.soft02007M}
}

@article{Bempong_Manful_2020,
   title={A high-resolution view of the jets in 3C 465},
   volume={496},
   ISSN={1365-2966},
   url={http://dx.doi.org/10.1093/mnras/staa1471},
   DOI={10.1093/mnras/staa1471},
   number={1},
   journal={Monthly Notices of the Royal Astronomical Society},
   publisher={Oxford University Press (OUP)},
   author={Bempong-Manful, E and Hardcastle, M J and Birkinshaw, M and Laing, R A and Leahy, J P and Worrall, D M},
   year={2020},
   month=may, pages={676–688} }

@ARTICLE{Hardcastle2025,
       author = {{Hardcastle}, M.~J. and {Pierce}, J.~C.~S. and {Duncan}, K.~J. and {G{\"u}rkan}, G. and {Gong}, Y. and {Horton}, M.~A. and {Mingo}, B. and {R{\"o}ttgering}, H.~J.~A. and {Smith}, D.~J.~B.},
        title = "{Radio AGN selection in LoTSS DR2}",
      journal = {arXiv e-prints},
         year = 2025,
        month = apr,
          eid = {arXiv:2504.09303},
        pages = {arXiv:2504.09303},
          doi = {10.48550/arXiv.2504.09303},
archivePrefix = {arXiv},
       eprint = {2504.09303},
 primaryClass = {astro-ph.GA},
       adsurl = {https://ui.adsabs.harvard.edu/abs/2025arXiv250409303H}
}

@ARTICLE{Hardcastle2016,
       author = {{Hardcastle}, M.~J. and {Lenc}, E. and {Birkinshaw}, M. and {Croston}, J.~H. and {Goodger}, J.~L. and {Marshall}, H.~L. and {Perlman}, E.~S. and {Siemiginowska}, A. and {Stawarz}, {\L}. and {Worrall}, D.~M.},
        title = "{Deep Chandra observations of Pictor A}",
      journal = {\mnras},
         year = 2016,
        month = feb,
       volume = {455},
       number = {4},
        pages = {3526-3545},
          doi = {10.1093/mnras/stv2553},
archivePrefix = {arXiv},
       eprint = {1510.08392},
 primaryClass = {astro-ph.HE},
       adsurl = {https://ui.adsabs.harvard.edu/abs/2016MNRAS.455.3526H}
}

@ARTICLE{GopalKrishna2024,
       author = {{Gopal-Krishna} and {Biermann}, Peter L.},
        title = "{Collimated synchrotron threads in wide-angle-tail radio galaxies: cosmic thunderbolts?}",
      journal = {\mnras},
         year = 2024,
        month = mar,
       volume = {529},
       number = {1},
        pages = {L135-L139},
          doi = {10.1093/mnrasl/slad191},
       adsurl = {https://ui.adsabs.harvard.edu/abs/2024MNRAS.529L.135G}
}

@ARTICLE{Koribalski2024,
       author = {{Koribalski}, B{\"a}rbel S. and {Duchesne}, Stefan W. and {Lenc}, Emil and {Venturi}, Tiziana and {Botteon}, Andrea and {Shabala}, Stanislav S. and {Vernstrom}, Tessa and {Carretti}, Ettore and {Norris}, Ray P. and {Anderson}, Craig and {Hopkins}, Andrew M. and {Riseley}, C.~J. and {Gupta}, Nikhel and {Velovi{\'c}}, Velibor},
        title = "{ASKAP reveals the radio tail structure of the Corkscrew Galaxy shaped by its passage through the Abell 3627 cluster}",
      journal = {\mnras},
         year = 2024,
        month = sep,
       volume = {533},
       number = {1},
        pages = {608-620},
          doi = {10.1093/mnras/stae1838},
archivePrefix = {arXiv},
       eprint = {2405.04374},
 primaryClass = {astro-ph.GA},
       adsurl = {https://ui.adsabs.harvard.edu/abs/2024MNRAS.533..608K}
}

@ARTICLE{Ledlow1996,
       author = {{Ledlow}, Michael J. and {Owen}, Frazer N.},
        title = "{20 CM VLA Survey of Abell Clusters of Galaxies. VI. Radio/Optical Luminosity Functions}",
      journal = {\aj},
         year = 1996,
        month = jul,
       volume = {112},
        pages = {9},
          doi = {10.1086/117985},
archivePrefix = {arXiv},
       eprint = {astro-ph/9607014},
 primaryClass = {astro-ph},
       adsurl = {https://ui.adsabs.harvard.edu/abs/1996AJ....112....9L}
}

@ARTICLE{Gizani2003,
       author = {{Gizani}, Nectaria A.~B. and {Leahy}, J.~P.},
        title = "{A multiband study of Hercules A - II. Multifrequency Very Large Array imaging}",
      journal = {\mnras},
         year = 2003,
        month = jun,
       volume = {342},
       number = {2},
        pages = {399-421},
          doi = {10.1046/j.1365-8711.2003.06469.x},
archivePrefix = {arXiv},
       eprint = {astro-ph/0305600},
 primaryClass = {astro-ph},
       adsurl = {https://ui.adsabs.harvard.edu/abs/2003MNRAS.342..399G}
}

@ARTICLE{Ramatsoku2020,
       author = {{Ramatsoku}, M. and {Murgia}, M. and {Vacca}, V. and {Serra}, P. and {Makhathini}, S. and {Govoni}, F. and {Smirnov}, O. and {Andati}, L.~A.~L. and {de Blok}, E. and {J{\'o}zsa}, G.~I.~G. and {Kamphuis}, P. and {Kleiner}, D. and {Maccagni}, F.~M. and {Moln{\'a}r}, D. Cs. and {Ramaila}, A.~J.~T. and {Thorat}, K. and {White}, S.~V.},
        title = "{Collimated synchrotron threads linking the radio lobes of ESO 137-006}",
      journal = {\aap},
         year = 2020,
        month = apr,
       volume = {636},
          eid = {L1},
        pages = {L1},
          doi = {10.1051/0004-6361/202037800},
archivePrefix = {arXiv},
       eprint = {2004.03203},
 primaryClass = {astro-ph.GA},
       adsurl = {https://ui.adsabs.harvard.edu/abs/2020A&A...636L...1R}
}

@ARTICLE{Heyvaerts1989,
       author = {{Heyvaerts}, Jean and {Norman}, Colin},
        title = "{The Collimation of Magnetized Winds}",
      journal = {\apj},
         year = 1989,
        month = dec,
       volume = {347},
        pages = {1055},
          doi = {10.1086/168195},
       adsurl = {https://ui.adsabs.harvard.edu/abs/1989ApJ...347.1055H}
}

@ARTICLE{Shimwell2017,
       author = {{Shimwell}, T.~W. and {R{\"o}ttgering}, H.~J.~A. and {Best}, P.~N. and {Williams}, W.~L. and {Dijkema}, T.~J. and {de Gasperin}, F. and {Hardcastle}, M.~J. and {Heald}, G.~H. and {Hoang}, D.~N. and {Horneffer}, A. and {Intema}, H. and {Mahony}, E.~K. and {Mandal}, S. and {Mechev}, A.~P. and {Morabito}, L. and {Oonk}, J.~B.~R. and {Rafferty}, D. and {Retana-Montenegro}, E. and {Sabater}, J. and {Tasse}, C. and {van Weeren}, R.~J. and {Br{\"u}ggen}, M. and {Brunetti}, G. and {Chy{\.z}y}, K.~T. and {Conway}, J.~E. and {Haverkorn}, M. and {Jackson}, N. and {Jarvis}, M.~J. and {McKean}, J.~P. and {Miley}, G.~K. and {Morganti}, R. and {White}, G.~J. and {Wise}, M.~W. and {van Bemmel}, I.~M. and {Beck}, R. and {Brienza}, M. and {Bonafede}, A. and {Calistro Rivera}, G. and {Cassano}, R. and {Clarke}, A.~O. and {Cseh}, D. and {Deller}, A. and {Drabent}, A. and {van Driel}, W. and {Engels}, D. and {Falcke}, H. and {Ferrari}, C. and {Fr{\"o}hlich}, S. and {Garrett}, M.~A. and {Harwood}, J.~J. and {Heesen}, V. and {Hoeft}, M. and {Horellou}, C. and {Israel}, F.~P. and {Kapi{\'n}ska}, A.~D. and {Kunert-Bajraszewska}, M. and {McKay}, D.~J. and {Mohan}, N.~R. and {Orr{\'u}}, E. and {Pizzo}, R.~F. and {Prandoni}, I. and {Schwarz}, D.~J. and {Shulevski}, A. and {Sipior}, M. and {Smith}, D.~J.~B. and {Sridhar}, S.~S. and {Steinmetz}, M. and {Stroe}, A. and {Varenius}, E. and {van der Werf}, P.~P. and {Zensus}, J.~A. and {Zwart}, J.~T.~L.},
        title = "{The LOFAR Two-metre Sky Survey. I. Survey description and preliminary data release}",
      journal = {\aap},
         year = 2017,
        month = feb,
       volume = {598},
          eid = {A104},
        pages = {A104},
          doi = {10.1051/0004-6361/201629313},
archivePrefix = {arXiv},
       eprint = {1611.02700},
 primaryClass = {astro-ph.IM},
       adsurl = {https://ui.adsabs.harvard.edu/abs/2017A&A...598A.104S}
}

@ARTICLE{Hardcastle2020,
       author = {{Hardcastle}, M.~J. and {Croston}, J.~H.},
        title = "{Radio galaxies and feedback from AGN jets}",
      journal = {\nar},
         year = 2020,
        month = jun,
       volume = {88},
          eid = {101539},
        pages = {101539},
          doi = {10.1016/j.newar.2020.101539},
archivePrefix = {arXiv},
       eprint = {2003.06137},
 primaryClass = {astro-ph.HE},
       adsurl = {https://ui.adsabs.harvard.edu/abs/2020NewAR..8801539H}
}

@ARTICLE{Perucho2007,
       author = {{Perucho}, M. and {Mart{\'\i}}, J.~M.},
        title = "{A numerical simulation of the evolution and fate of a Fanaroff-Riley type I jet. The case of 3C 31}",
      journal = {\mnras},
         year = 2007,
        month = dec,
       volume = {382},
       number = {2},
        pages = {526-542},
          doi = {10.1111/j.1365-2966.2007.12454.x},
archivePrefix = {arXiv},
       eprint = {0709.1784},
 primaryClass = {astro-ph},
       adsurl = {https://ui.adsabs.harvard.edu/abs/2007MNRAS.382..526P}
}

@ARTICLE{Mingo2014,
       author = {{Mingo}, B. and {Hardcastle}, M.~J. and {Croston}, J.~H. and {Dicken}, D. and {Evans}, D.~A. and {Morganti}, R. and {Tadhunter}, C.},
        title = "{An X-ray survey of the 2 Jy sample - I. Is there an accretion mode dichotomy in radio-loud AGN?}",
      journal = {\mnras},
         year = 2014,
        month = may,
       volume = {440},
       number = {1},
        pages = {269-297},
          doi = {10.1093/mnras/stu263},
archivePrefix = {arXiv},
       eprint = {1402.1770},
 primaryClass = {astro-ph.GA},
       adsurl = {https://ui.adsabs.harvard.edu/abs/2014MNRAS.440..269M}
}

@ARTICLE{Govoni_2019,
       author = {{Govoni}, F. and {Orr{\`u}}, E. and {Bonafede}, A. and {Iacobelli}, M. and {Paladino}, R. and {Vazza}, F. and {Murgia}, M. and {Vacca}, V. and {Giovannini}, G. and {Feretti}, L. and {Loi}, F. and {Bernardi}, G. and {Ferrari}, C. and {Pizzo}, R.~F. and {Gheller}, C. and {Manti}, S. and {Br{\"u}ggen}, M. and {Brunetti}, G. and {Cassano}, R. and {de Gasperin}, F. and {En{\ss}lin}, T.~A. and {Hoeft}, M. and {Horellou}, C. and {Junklewitz}, H. and {R{\"o}ttgering}, H.~J.~A. and {Scaife}, A.~M.~M. and {Shimwell}, T.~W. and {van Weeren}, R.~J. and {Wise}, M.},
        title = "{A radio ridge connecting two galaxy clusters in a filament of the cosmic web}",
      journal = {Science},
         year = 2019,
        month = jun,
       volume = {364},
       number = {6444},
        pages = {981-984},
          doi = {10.1126/science.aat7500},
archivePrefix = {arXiv},
       eprint = {1906.07584},
 primaryClass = {astro-ph.GA},
       adsurl = {https://ui.adsabs.harvard.edu/abs/2019Sci...364..981G}
}

@ARTICLE{van_Weeren_2019,
       author = {{van Weeren}, R.~J. and {de Gasperin}, F. and {Akamatsu}, H. and {Br{\"u}ggen}, M. and {Feretti}, L. and {Kang}, H. and {Stroe}, A. and {Zandanel}, F.},
        title = "{Diffuse Radio Emission from Galaxy Clusters}",
      journal = {\ssr},
         year = 2019,
        month = feb,
       volume = {215},
       number = {1},
          eid = {16},
        pages = {16},
          doi = {10.1007/s11214-019-0584-z},
archivePrefix = {arXiv},
       eprint = {1901.04496},
 primaryClass = {astro-ph.HE},
       adsurl = {https://ui.adsabs.harvard.edu/abs/2019SSRv..215...16V}
}

@ARTICLE{Anderson2018,
       author = {{Anderson}, Craig S. and {Heald}, George and {O'Sullivan}, Shane P. and {Bunton}, John D. and {Carretti}, Ettore and {Chippendale}, Aaron P. and {Collier}, Jordan D. and {Farnes}, Jamie S. and {Gaensler}, Bryan M. and {Harvey-Smith}, Lisa and {Koribalski}, B{\"a}rbel S. and {Landecker}, Tom L. and {Lenc}, Emil and {McClure-Griffiths}, Naomi M. and {Mitchell}, Daniel and {Rudnick}, Lawrence and {West}, Jennifer},
        title = "{The Extraordinary Linear Polarisation Structure of the Southern Centaurus A Lobe Revealed by ASKAP}",
      journal = {Galaxies},
         year = 2018,
        month = nov,
       volume = {6},
       number = {4},
          eid = {127},
        pages = {127},
          doi = {10.3390/galaxies6040127},
archivePrefix = {arXiv},
       eprint = {1811.11760},
 primaryClass = {astro-ph.GA},
       adsurl = {https://ui.adsabs.harvard.edu/abs/2018Galax...6..127A}

}

@ARTICLE{Perley1984,
       author = {{Perley}, R.~A. and {Dreher}, J.~W. and {Cowan}, J.~J.},
        title = "{The jet and filaments in Cygnus A.}",
      journal = {\apjl},
         year = 1984,
        month = oct,
       volume = {285},
        pages = {L35-L38},
          doi = {10.1086/184360},
       adsurl = {https://ui.adsabs.harvard.edu/abs/1984ApJ...285L..35P}
}

@ARTICLE{Brunetti2014,
       author = {{Brunetti}, Gianfranco and {Jones}, Thomas W.},
        title = "{Cosmic Rays in Galaxy Clusters and Their Nonthermal Emission}",
      journal = {International Journal of Modern Physics D},
         year = 2014,
        month = mar,
       volume = {23},
       number = {4},
          eid = {1430007-98},
        pages = {1430007-98},
          doi = {10.1142/S0218271814300079},
archivePrefix = {arXiv},
       eprint = {1401.7519},
 primaryClass = {astro-ph.CO},
       adsurl = {https://ui.adsabs.harvard.edu/abs/2014IJMPD..2330007B}
}

@article{ Böhringer2014,
       author = {{B{\"o}hringer}, Hans and {Chon}, Gayoung and {Collins}, Chris A.},
        title = "{The extended ROSAT-ESO Flux Limited X-ray Galaxy Cluster Survey (REFLEX II). IV. X-ray luminosity function and first constraints on cosmological parameters}",
      journal = {\aap},
         year = 2014,
        month = oct,
       volume = {570},
          eid = {A31},
        pages = {A31},
          doi = {10.1051/0004-6361/201323155},
archivePrefix = {arXiv},
       eprint = {1403.2927},
 primaryClass = {astro-ph.CO},
       adsurl = {https://ui.adsabs.harvard.edu/abs/2014A&A...570A..31B}
}

@ARTICLE{Snowden1994,
       author = {{Snowden}, S.~L. and {McCammon}, D. and {Burrows}, D.~N. and {Mendenhall}, J.~A.},
        title = "{Analysis Procedures for ROSAT XRT/PSPC Observations of Extended Objects and the Diffuse Background}",
      journal = {\apj},
         year = 1994,
        month = apr,
       volume = {424},
        pages = {714},
          doi = {10.1086/173925},
       adsurl = {https://ui.adsabs.harvard.edu/abs/1994ApJ...424..714S}
}

@ARTICLE{Shimwell2022,
       author = {{Shimwell}, T.~W. and {Hardcastle}, M.~J. and {Tasse}, C. and {Best}, P.~N. and {R{\"o}ttgering}, H.~J.~A. and {Williams}, W.~L. and {Botteon}, A. and {Drabent}, A. and {Mechev}, A. and {Shulevski}, A. and {van Weeren}, R.~J. and {Bester}, L. and {Br{\"u}ggen}, M. and {Brunetti}, G. and {Callingham}, J.~R. and {Chy{\.z}y}, K.~T. and {Conway}, J.~E. and {Dijkema}, T.~J. and {Duncan}, K. and {de Gasperin}, F. and {Hale}, C.~L. and {Haverkorn}, M. and {Hugo}, B. and {Jackson}, N. and {Mevius}, M. and {Miley}, G.~K. and {Morabito}, L.~K. and {Morganti}, R. and {Offringa}, A. and {Oonk}, J.~B.~R. and {Rafferty}, D. and {Sabater}, J. and {Smith}, D.~J.~B. and {Schwarz}, D.~J. and {Smirnov}, O. and {O'Sullivan}, S.~P. and {Vedantham}, H. and {White}, G.~J. and {Albert}, J.~G. and {Alegre}, L. and {Asabere}, B. and {Bacon}, D.~J. and {Bonafede}, A. and {Bonnassieux}, E. and {Brienza}, M. and {Bilicki}, M. and {Bonato}, M. and {Calistro Rivera}, G. and {Cassano}, R. and {Cochrane}, R. and {Croston}, J.~H. and {Cuciti}, V. and {Dallacasa}, D. and {Danezi}, A. and {Dettmar}, R.~J. and {Di Gennaro}, G. and {Edler}, H.~W. and {En{\ss}lin}, T.~A. and {Emig}, K.~L. and {Franzen}, T.~M.~O. and {Garc{\'\i}a-Vergara}, C. and {Grange}, Y.~G. and {G{\"u}rkan}, G. and {Hajduk}, M. and {Heald}, G. and {Heesen}, V. and {Hoang}, D.~N. and {Hoeft}, M. and {Horellou}, C. and {Iacobelli}, M. and {Jamrozy}, M. and {Jeli{\'c}}, V. and {Kondapally}, R. and {Kukreti}, P. and {Kunert-Bajraszewska}, M. and {Magliocchetti}, M. and {Mahatma}, V. and {Ma{\l}ek}, K. and {Mandal}, S. and {Massaro}, F. and {Meyer-Zhao}, Z. and {Mingo}, B. and {Mostert}, R.~I.~J. and {Nair}, D.~G. and {Nakoneczny}, S.~J. and {Nikiel-Wroczy{\'n}ski}, B. and {Orr{\'u}}, E. and {Pajdosz-{\'S}mierciak}, U. and {Pasini}, T. and {Prandoni}, I. and {van Piggelen}, H.~E. and {Rajpurohit}, K. and {Retana-Montenegro}, E. and {Riseley}, C.~J. and {Rowlinson}, A. and {Saxena}, A. and {Schrijvers}, C. and {Sweijen}, F. and {Siewert}, T.~M. and {Timmerman}, R. and {Vaccari}, M. and {Vink}, J. and {West}, J.~L. and {Wo{\l}owska}, A. and {Zhang}, X. and {Zheng}, J.},
        title = "{The LOFAR Two-metre Sky Survey. V. Second data release}",
      journal = {\aap},
         year = 2022,
        month = mar,
       volume = {659},
          eid = {A1},
        pages = {A1},
          doi = {10.1051/0004-6361/202142484},
archivePrefix = {arXiv},
       eprint = {2202.11733},
 primaryClass = {astro-ph.GA},
       adsurl = {https://ui.adsabs.harvard.edu/abs/2022A&A...659A...1S}
}

@ARTICLE{Hardcastle-2018,
       author = {{Hardcastle}, M.~J.},
        title = "{A simulation-based analytic model of radio galaxies}",
      journal = {\mnras},
         year = 2018,
        month = apr,
       volume = {475},
       number = {2},
        pages = {2768-2786},
          doi = {10.1093/mnras/stx3358},
archivePrefix = {arXiv},
       eprint = {1801.00667},
 primaryClass = {astro-ph.HE},
       adsurl = {https://ui.adsabs.harvard.edu/abs/2018MNRAS.475.2768H}
}

@ARTICLE{van-Haarlem2013,
       author = {{van Haarlem}, M.~P. and {Wise}, M.~W. and {Gunst}, A.~W. and {Heald}, G. and {McKean}, J.~P. and {Hessels}, J.~W.~T. and {de Bruyn}, A.~G. and {Nijboer}, R. and {Swinbank}, J. and {Fallows}, R. and {Brentjens}, M. and {Nelles}, A. and {Beck}, R. and {Falcke}, H. and {Fender}, R. and {H{\"o}randel}, J. and {Koopmans}, L.~V.~E. and {Mann}, G. and {Miley}, G. and {R{\"o}ttgering}, H. and {Stappers}, B.~W. and {Wijers}, R.~A.~M.~J. and {Zaroubi}, S. and {van den Akker}, M. and {Alexov}, A. and {Anderson}, J. and {Anderson}, K. and {van Ardenne}, A. and {Arts}, M. and {Asgekar}, A. and {Avruch}, I.~M. and {Batejat}, F. and {B{\"a}hren}, L. and {Bell}, M.~E. and {Bell}, M.~R. and {van Bemmel}, I. and {Bennema}, P. and {Bentum}, M.~J. and {Bernardi}, G. and {Best}, P. and {B{\^\i}rzan}, L. and {Bonafede}, A. and {Boonstra}, A. -J. and {Braun}, R. and {Bregman}, J. and {Breitling}, F. and {van de Brink}, R.~H. and {Broderick}, J. and {Broekema}, P.~C. and {Brouw}, W.~N. and {Br{\"u}ggen}, M. and {Butcher}, H.~R. and {van Cappellen}, W. and {Ciardi}, B. and {Coenen}, T. and {Conway}, J. and {Coolen}, A. and {Corstanje}, A. and {Damstra}, S. and {Davies}, O. and {Deller}, A.~T. and {Dettmar}, R. -J. and {van Diepen}, G. and {Dijkstra}, K. and {Donker}, P. and {Doorduin}, A. and {Dromer}, J. and {Drost}, M. and {van Duin}, A. and {Eisl{\"o}ffel}, J. and {van Enst}, J. and {Ferrari}, C. and {Frieswijk}, W. and {Gankema}, H. and {Garrett}, M.~A. and {de Gasperin}, F. and {Gerbers}, M. and {de Geus}, E. and {Grie{\ss}meier}, J. -M. and {Grit}, T. and {Gruppen}, P. and {Hamaker}, J.~P. and {Hassall}, T. and {Hoeft}, M. and {Holties}, H.~A. and {Horneffer}, A. and {van der Horst}, A. and {van Houwelingen}, A. and {Huijgen}, A. and {Iacobelli}, M. and {Intema}, H. and {Jackson}, N. and {Jelic}, V. and {de Jong}, A. and {Juette}, E. and {Kant}, D. and {Karastergiou}, A. and {Koers}, A. and {Kollen}, H. and {Kondratiev}, V.~I. and {Kooistra}, E. and {Koopman}, Y. and {Koster}, A. and {Kuniyoshi}, M. and {Kramer}, M. and {Kuper}, G. and {Lambropoulos}, P. and {Law}, C. and {van Leeuwen}, J. and {Lemaitre}, J. and {Loose}, M. and {Maat}, P. and {Macario}, G. and {Markoff}, S. and {Masters}, J. and {McFadden}, R.~A. and {McKay-Bukowski}, D. and {Meijering}, H. and {Meulman}, H. and {Mevius}, M. and {Middelberg}, E. and {Millenaar}, R. and {Miller-Jones}, J.~C.~A. and {Mohan}, R.~N. and {Mol}, J.~D. and {Morawietz}, J. and {Morganti}, R. and {Mulcahy}, D.~D. and {Mulder}, E. and {Munk}, H. and {Nieuwenhuis}, L. and {van Nieuwpoort}, R. and {Noordam}, J.~E. and {Norden}, M. and {Noutsos}, A. and {Offringa}, A.~R. and {Olofsson}, H. and {Omar}, A. and {Orr{\'u}}, E. and {Overeem}, R. and {Paas}, H. and {Pandey-Pommier}, M. and {Pandey}, V.~N. and {Pizzo}, R. and {Polatidis}, A. and {Rafferty}, D. and {Rawlings}, S. and {Reich}, W. and {de Reijer}, J. -P. and {Reitsma}, J. and {Renting}, G.~A. and {Riemers}, P. and {Rol}, E. and {Romein}, J.~W. and {Roosjen}, J. and {Ruiter}, M. and {Scaife}, A. and {van der Schaaf}, K. and {Scheers}, B. and {Schellart}, P. and {Schoenmakers}, A. and {Schoonderbeek}, G. and {Serylak}, M. and {Shulevski}, A. and {Sluman}, J. and {Smirnov}, O. and {Sobey}, C. and {Spreeuw}, H. and {Steinmetz}, M. and {Sterks}, C.~G.~M. and {Stiepel}, H. -J. and {Stuurwold}, K. and {Tagger}, M. and {Tang}, Y. and {Tasse}, C. and {Thomas}, I. and {Thoudam}, S. and {Toribio}, M.~C. and {van der Tol}, B. and {Usov}, O. and {van Veelen}, M. and {van der Veen}, A. -J. and {ter Veen}, S. and {Verbiest}, J.~P.~W. and {Vermeulen}, R. and {Vermaas}, N. and {Vocks}, C. and {Vogt}, C. and {de Vos}, M. and {van der Wal}, E. and {van Weeren}, R. and {Weggemans}, H. and {Weltevrede}, P. and {White}, S. and {Wijnholds}, S.~J. and {Wilhelmsson}, T. and {Wucknitz}, O. and {Yatawatta}, S. and {Zarka}, P. and {Zensus}, A. and {van Zwieten}, J.},
        title = "{LOFAR: The LOw-Frequency ARray}",
      journal = {\aap},
         year = 2013,
        month = aug,
       volume = {556},
          eid = {A2},
        pages = {A2},
          doi = {10.1051/0004-6361/201220873},
archivePrefix = {arXiv},
       eprint = {1305.3550},
 primaryClass = {astro-ph.IM},
       adsurl = {https://ui.adsabs.harvard.edu/abs/2013A&A...556A...2V}
}

@book{Rybicki1979,
  title={Radiative processes in astrophysics},
  author={Rybicki, George B and Lightman, Alan P},
  year={2024},
  publisher={John Wiley \& Sons}
}

@ARTICLE{Arnaud2010,
       author = {{Arnaud}, M. and {Pratt}, G.~W. and {Piffaretti}, R. and {B{\"o}hringer}, H. and {Croston}, J.~H. and {Pointecouteau}, E.},
        title = "{The universal galaxy cluster pressure profile from a representative sample of nearby systems (REXCESS) and the Y$_{SZ}$ - M$_{500}$ relation}",
      journal = {\aap},
         year = 2010,
        month = jul,
       volume = {517},
          eid = {A92},
        pages = {A92},
          doi = {10.1051/0004-6361/200913416},
archivePrefix = {arXiv},
       eprint = {0910.1234},
 primaryClass = {astro-ph.CO},
       adsurl = {https://ui.adsabs.harvard.edu/abs/2010A&A...517A..92A}
}

\end{document}